\documentclass[12pt]{article}
\usepackage{amsmath,amssymb}
\usepackage{epsfig} 
\usepackage{cite}

\def\m@th{\mathsurround=0pt }
\def\leftrightarrowfill{$\m@th \mathord\leftarrow \mkern-6mu
	\cleaders\hbox{$\mkern-2mu \mathord- \mkern-2mu$}\hfill
	\mkern-6mu \mathord\rightarrow$}

\def\overleftrightarrow#1{\vbox{\ialign{##\crcr
	\leftrightarrowfill\crcr\noalign{\kern-1pt\nointerlineskip}
	$\hfil\displaystyle{#1}\hfil$\crcr}}}

\newcommand{\tr}{{\rm Tr}}
\newcommand{\be}{\begin{equation}}
\newcommand{\ee}{\end{equation}}

\def\shat{\ifmmode \hat{s}\else $\hat{s}$\fi}

\newcommand{\newc}{\newcommand}

\newc{\gsim}{\lower.7ex\hbox{$\;\stackrel{\textstyle>}{\sim}\;$}}
\newc{\lsim}{\lower.7ex\hbox{$\;\stackrel{\textstyle<}{\sim}\;$}}
\newc{\ie}{{\it i.e.}}
\newc{\etal}{{\it et al.}}
\newc{\mev}{\hbox{\rm\,MeV}}
\newc{\gev}{\hbox{\rm\,GeV}}
\newc{\tev}{\hbox{\rm\,TeV}}
\newc{\xpb}{\hbox{\rm\, pb}}
\newc{\xfb}{\hbox{\rm\, fb}}

\newc{\G}{{\cal G}}
\newc{\h}{{\cal H}}
\newc{\D}{{\cal D}}
\newc{\E}{{\cal E}}

\newc{\mtop}{m_t}
\newc{\mbot}{m_b}
\newc{\mz}{M_Z}
\newc{\mw}{M_W}
\newc{\alphasmz}{\alpha_s(M_Z)}
\newc{\swsq}{\sin^2\theta_W}
\newc{\cwsq}{\cos^2\theta_W}
\newc{\tw}{\tan\theta_W}
\newc{\cw}{\cos\theta_W}
\newc{\sw}{\sin\theta_W}
\newc{\BR}{\hbox{\rm BR}}
\newc{\zbb}{Z\to b\bar}
\newc{\Gb}{\Gamma (Z\to b\bar b)}
\newc{\Gh}{\Gamma (Z\to \hbox{\rm hadrons})}
\newc{\sgn}{\mbox{sgn}}

\def\eq#1{eq.~(\ref{#1})}

\def\vev#1{\langle {#1} \rangle}

\def\gsm{g_{\scriptscriptstyle {SM}}}
\def\asm{\alpha_{\scriptscriptstyle {SM}}}
\def\grho{g_{\rho}}

\newcounter{mysubequation}[equation]

\newcommand{\GeV}{\,\mathrm{GeV}}

\def\beq{\begin{equation}}
\def\eeq{\end{equation}}
\def\bea{\begin{eqnarray}}
\def\eea{\end{eqnarray}}
\def\slashchar#1{\setbox0=\hbox{$#1$}           
   \dimen0=\wd0                                 
   \setbox1=\hbox{/} \dimen1=\wd1               
   \ifdim\dimen0>\dimen1                        
      \rlap{\hbox to \dimen0{\hfil/\hfil}}      
      #1                                        
   \else                                        
      \rlap{\hbox to \dimen1{\hfil$#1$\hfil}}   
      /                                         
   \fi}                                         %
\catcode`@=11
\long\def\@caption#1[#2]#3{\par\addcontentsline{\csname
  ext@#1\endcsname}{#1}{\protect\numberline{\csname
  the#1\endcsname}{\ignorespaces #2}}\begingroup
    \small
    \@parboxrestore
    \@makecaption{\csname fnum@#1\endcsname}{\ignorespaces #3}\par
  \endgroup}
\catcode`@=12

\begin{document}

\baselineskip=18pt

\setcounter{footnote}{0}
\setcounter{figure}{0}
\setcounter{table}{0}

\begin{titlepage}
\begin{flushright}
CERN-PH-TH/2007--47
\end{flushright}
\vspace{.3in}

\begin{center}
{\Large \bf The Strongly-Interacting Light Higgs}

\vspace{0.5cm}

{\bf G. F. Giudice$^{a}$, C. Grojean$^{a,b}$, A. Pomarol$^{c}$, R. Rattazzi$^{a,d}$}

\vspace{.5cm}

\centerline{$^{a}${\it CERN, Theory Division, CH--1211 Geneva 23, Switzerland}}
\centerline{$^{b}${\it Service de Physique Th\'eorique, CEA Saclay, F91191 Gif--sur--Yvette,
France}}
\centerline{$^{c}${\it IFAE, Universitat Aut\`onoma de Barcelona, 08193 Bellaterra, Barcelona, Spain}}
\centerline{$^{d}${\it Institut de Th\'eorie des Ph\'enom\`enes Physiques, EPFL,  CH--1015 Lausanne, Switzerland}}

\end{center}
\vspace{.8cm}

\begin{abstract}
\medskip
\noindent
We develop a simple description of models where electroweak symmetry breaking is triggered by a light composite Higgs, which emerges from a strongly-interacting sector as a pseudo-Goldstone boson. Two parameters broadly characterize these models: $m_\rho$, the mass scale of the new resonances and $g_\rho$, their coupling. An effective low-energy Lagrangian approach proves to be useful for LHC and ILC phenomenology below the scale $m_\rho$. We identify two classes of operators: those that are genuinely sensitive to the new strong force and those that are sensitive to the spectrum of the resonances only. Phenomenological prospects for the LHC and the ILC include the study of high-energy longitudinal vector boson scattering, strong double-Higgs production and anomalous Higgs couplings.  We finally discuss the possibility that the top quark could also be a composite object of the strong sector.
\end{abstract}

\bigskip
\bigskip

\end{titlepage}


\section{Introduction}
\label{intro}

The main goal of the LHC is to unveil the mechanism of electroweak symmetry breaking. A crucial issue that experiments should be able to settle is whether the dynamics responsible for symmetry breaking is weakly or strongly coupled. LEP1 has provided us with convincing indications in favor of weakly-coupled dynamics. Indeed, the good agreement of precision measurements with the Standard Model (SM) predictions showed that the new dynamics cannot significantly influence the properties of the $Z$ boson, ruling out, for instance, the simplest forms of technicolor models, which were viewed as the prototypes of a strongly-interacting electroweak sector. Moreover, the best agreement between experiments and theory was obtained for a light Higgs,  corresponding to a weakly-coupled Higgs self-interaction. Finally, supersymmetry, which appeared to be the most realistic realization of a light Higgs with mass stabilized under quantum corrections, received a further boost by the LEP1 measurements of gauge coupling constants, found to be in accord with supersymmetric unification.

The situation has swayed back after the LEP2 results. The lack of discovery of a Higgs boson below 114~GeV or of any new states has forced supersymmetry into fine-tuning territory, partially undermining its original motivation. Moreover, new theoretical developments, mostly influenced by extra dimensions and by the connection between strongly-interacting gauge theories and gravity on warped geometries, have led the way to the construction of new models of electroweak symmetry breaking~\cite{Arkani-Hamed:2001nc, Arkani-Hamed:2002qy, gaugehiggs, higgsless, Contino:2003ve, Agashe:2004rs}. Still, the complete replacement of the Higgs sector with strongly-interacting dynamics seemed hard to implement, mostly because of constraints from electroweak data. A more promising approach is to keep the Higgs boson as an effective field arising from new dynamics~\cite{georgikaplan,othercompositeHiggs} which becomes strong at a scale not much larger than the Fermi scale. There has been various attempts to realize such scenario, including the Little Higgs~\cite{Arkani-Hamed:2002qy}, Holographic Higgs as Goldstone bosons~\cite{Contino:2003ve, Agashe:2004rs} or not~\cite{Agashe:2003zs}, and other variations.

In this paper we want to study the general properties and the phenomenology of scenarios in which a light Higgs is associated with strong dynamics at a higher scale, focusing on features that are quite independent of the particular model realization. We will refer to this scenario as to the Strongly-Interacting Light Higgs (SILH). Of course, in many specific models, the best experimental signals will be provided by direct production of new states, while here we concentrate on deviations from SM properties in Higgs and longitudinal gauge boson processes. Still, we believe that our model-independent approach is useful. The tests we propose here on Higgs and gauge-boson interactions will help, in case of new discoveries, to establish if the new particles indeed belong to a strongly-interacting sector ultimately responsible for electroweak symmetry breaking. If no new states are observed, or if the resonances are too broad to be identified, then our tests can be used to investigate whether the Higgs is weakly coupled or is an effective particle emerging from a strongly-interacting sector, whose discovery has been barely missed by direct searches at the LHC.  

This paper is organized as follows. In sect.~\ref{secstruc}, we define the SILH and construct the low-energy effective theory that describes its interactions with the SM fields. In sect.~\ref{secmod}, we discuss how this effective Lagrangian is related to specific models previously proposed in the literature, like the Holographic Higgs and the Little Higgs. Then we describe in sect.~\ref{secphen}, how the SILH can be tested in collider experiments. In sect.~\ref{sectop}, we extend our analysis to the case of a composite top quark and finally we summarize our results and draw our conclusions in sect.~\ref{secconc}.

\section{The structure of SILH}
\label{secstruc}

\subsection{Definition of SILH}

The structure of the theories we want to consider is the  following. In addition to the vector bosons and fermions
of the SM, there exists a new  sector responsible for EW symmetry breaking, which is broadly characterized by two parameters, a coupling $g_\rho$ and a scale $m_\rho$ describing the mass of heavy physical states. Collectively indicating by $\gsm$ the SM gauge
and Yukawa couplings (basically the weak gauge coupling and the top quark Yukawa), we assume
$\gsm \lsim \grho \lsim 4 \pi$. The upper bound on $\grho$
ensures that the loop expansion parameter
$\sim (g_\rho/4\pi)^2$ is less than unity,  while the limit $g_\rho \sim 4 \pi$ corresponds to a maximally strongly-coupled theory in the spirit of naive dimensional analysis  (NDA)~\cite{nda}. Because of the first
inequality,  by a slight abuse of language, we shall refer to the new sector as ``the strong sector''.
The Higgs multiplet is assumed to belong to the 
strong sector. The SM vector bosons and fermions are weakly coupled to the strong sector by means
of the $SU(3)\times SU(2)\times U(1)_Y$ gauge coupling and by means of proto-Yukawa interactions,
namely interactions that in the low-energy effective field theory will give rise to the SM Yukawas.

A second crucial assumption we are going to make is that in the limit $\gsm=0$, $\grho\not=0$ the Higgs doublet $H$ is an exact Goldstone boson, living in the $\G/\h$ coset space of a spontaneously broken symmetry of the strong sector. Two minimal possibilities in which the complex Higgs doublet spans the whole coset space are $SU(3)/SU(2)\times U(1)$ and the custodially symmetric $SO(5)/SO(4)$.

The gauging of $SU(2)\times U(1)_Y$ and the non-zero Yukawas explicitly break
the Goldstone symmetry of the strong sector leading to terms in the (effective) action that are not invariant under the action of 
$\G$ on the coset space. In particular a mass term for the Higgs is generated at 1-loop. If the new dynamics is addressing the hierarchy problem, it should soften the sensitivity of the Higgs mass to short distances, that is to say below $1/m_\rho$. In interesting models, the Higgs mass parameter is thus expected to scale like $(\asm/4\pi) m_\rho^2$. Observation at the LHC of the new states with mass $m_\rho$ will be the key signature of the various realizations of SILH. Here, as stated in sect.~\ref{intro}, we are interested in the model-independent effects, which could be visible in processes involving the Higgs boson and/or longitudinal gauge bosons, and which would unmistakably reveal new physics in the electroweak breaking sector.

As we shall explain below, the $\sigma$-model scale $f$ is related to $g_\rho$ and $m_\rho$ by the equation
\beq
m_\rho =g_\rho f\, .
\label{usual}
\eeq
Fully strongly interacting theories, like QCD, correspond to $g_\rho \sim 4\pi$. In that case~\eq{usual} expresses the usual NDA relation between the pion decay constant $f$ and the mass scale of the QCD states.
On the other hand, the theories we are considering represent a ``weakly coupled" deformation of this QCD-like pattern. For $g_\rho<4\pi$,
the pure low-energy effective $\sigma$-model description  breaks down above a scale $m_\rho$, which is parametrically
lower than the scale $4\pi f$ where the $\sigma$-model would become strongly coupled. The coupling $g_\rho$ precisely measures how strong the coupling of the $\sigma$-model  can become before it  is  replaced by a more fundamental description. 
The simplest example of this possibility is represented by a linear $\sigma$-model UV completion of the $\G/\h$ non-linear theory. In this case the role of $m_\rho$ and $g_\rho^2$
is played respectively by the mass of the heavy scalar modes and by their quartic coupling.
For instance, in  the interesting case in which the Higgs complex
doublet spans $SO(5)/SO(4)$, a simple UV completion could consist of a real scalar $\Phi$
in the fundamental of $SO(5)$ and quartic potential~\footnote {Along the same lines we could even describe the SM in the limit of a heavy Higgs
boson, using an analogous $\sigma$-model involving only 3 (and not 4) Goldstone bosons, with $f=\vev{H}$,  $g_\rho =\sqrt{\lambda}$ (the quartic Higgs coupling), and the physical Higgs mass playing the role of $m_\rho$.}, $V=-m_\rho^2\Phi^2 +g_\rho^2 \Phi^4$.  
However, once the SM couplings are turned on, such a limited UV completion  fails to screen the quadratic corrections to the Higgs mass. 

A more interesting possibility  arises when the strong sector is composite so that the corrections to the Higgs mass are screened above the ``hadron" mass scale $m_\rho$. Moreover 
 if the underlying theory is a  large-$N$ gauge theory, we  also expect
 the hadrons to interact with a coupling
\beq
g_\rho =\frac{4\pi}{\sqrt{N}},
\label{largeN}
\eeq
which becomes weaker at large $N$.
This is also basically the picture that holds
in extra-dimensional constructions where the SM
is represented by a weakly-coupled boundary dynamics while the Higgs  sector is
part of a more strongly-coupled bulk dynamics. Examples of this type are the Holographic Goldstones~\cite{Agashe:2004rs} over a slice of AdS$_5$~\footnote{Notice that the AdS geometry only matters
for the extrapolation  to ultra-high scales. The same low-energy dynamics of ref.~\cite{Agashe:2004rs} could also be realized over flat space by turning on the suitable boundary terms, along the lines of ref.~\cite{Barbieri:2003pr} (although ref.~\cite{Barbieri:2003pr} focussed on the Higgsless limit).}. In these extra-dimensional realizations, the Kaluza--Klein mass and coupling play respectively  the role of $m_\rho$ and $g_\rho$,
 while the number of weakly-coupled Kaluza--Klein modes below the cut-off can be basically interpreted as $N$. To be explicit, consider a  5-dimensional (5D) gauge theory with 5D coupling $g_5$ compactified on a circle or orbifold of radius $R$. We have $m_\rho=1/R$ and $g_\rho^2\equiv g_5^2/(\pi R)$.  On the other hand, according to 5D NDA,
the physical cut-off of the model is $\Lambda \sim (4\pi)^2/g_5^2$. These relations then imply
\beq
\left (\frac{4\pi}{g_\rho}\right )^2=\Lambda \pi R\equiv N\, .
\eeq

Other models that basically fall into our class are Little Higgses~\cite{Arkani-Hamed:2001nc}. There, the  scale $m_\rho$ is represented
by the masses of the partners of top quark, electroweak vector bosons and Higgs, the states that soften
the quadratic correction to the Higgs mass. In Little-Higgs models there is more parameter freedom, and the
coupling $g_\rho$ is more accurately described by a set of couplings that can range from weak ($\sim\gsm$)
to strong ($\gg \gsm$).
Nonetheless we shall still find our simplified characterization very useful.

Summarizing, in several models of interest the electroweak-breaking sector corresponds to a ``deformation'' of a pure $\sigma$-model,  where weakly-coupled states appear below the naive cut-off $4\pi f$. It is useful
to focus on the simplest possibility where just one parameter, the coupling $g_\rho$ of the new states,
characterizes this ``deformation''~\footnote{Our simplified approach based on two parameters $(m_\rho, g_\rho) $ is close in spirit to recent studies of two (and three) site models \cite{Cheng:2006ht}. The phenomenological goal of our paper is however complementary to that refs. \cite{Cheng:2006ht}:  while those studies focus on the physics of the new heavy states, our paper focuses on the low-energy effects in Higgs and vector boson interactions.}.

\subsection{Constructing the effective action}
\label{secrules}

Under the assumptions of the class of theories defined above,
 we now want to derive the form of the most general effective Lagrangian for the SM + Higgs fields. Since we are assuming $g_\rho> \gsm$, it makes sense to focus first on the strong sector in the limit $\gsm=0$, and to turn on later the couplings of this sector to the SM vectors and fermions. Indicating by
$T^A$ and $T^a$  respectively the broken and the unbroken generators of the group $\G$, we parametrize the Goldstone field by the matrix
\beq
U= e^{i\Pi}\qquad\qquad \Pi \equiv \Pi^A T^A.
\label{Umatrix}
\eeq
   In addition we assume the strong sector
features a set of fields with mass of order $m_\rho$ , which we collectively indicate by $\Phi$. By our assumptions, the general form of the  action including quantum fluctuations from scales shorter than $1/m_\rho$ must be
\beq
{\cal L}= \frac{m_\rho^4}{g_\rho^2}\left[ {\cal L}^{(0)}(U,\Phi, \partial/m_\rho)+\frac{g_\rho^2}{(4\pi) ^2}{\cal L}^{(1)}(U,\Phi, \partial/m_\rho)+\frac{g_\rho^4}{(4\pi )^4} {\cal L}^{(2)}(U,\Phi, \partial/m_\rho)+\dots \right] .
\label{loopexpansion}
\eeq
In the action we have kept massive degrees of freedom ``integrated in'' for purposes that will become momentarily more clear. One can for instance check that the structure in~\eq{loopexpansion} is obtained in the compactification of a 5D gauge theory with the identification $m_\rho\equiv 1/R$
and $g_\rho^2=g_5^2/(\pi R)$ (provided the power divergent loops  are computed by NDA, while the log-divergent and finite pieces automatically satisfy the above structure). Moreover this same structure characterizes the effective field-theory description of the string. For instance in type I compactified
on a $T^6$ of radius $\sim 1/M_s$, we can make the identifications: $M_s=m_\rho$ and $g_s= g^2_\rho/2\pi$.

In order to  get the truly low-energy effective action we should then integrate out the $\Phi$'s and also include the quantum fluctuations at scales below $m_\rho$.
 If the structure of the terms in~\eq{loopexpansion} is the most general one, in particular if terms of all orders in
derivatives  appear already in the classical Lagrangian ${\cal L}^{(0)}$,
then the presence or absence of the $\Phi$'s has no impact
on the low-energy theory. We shall first concentrate on this case.
Later we shall discuss the
 more realistic situation where the classical Lagrangian involves at most two derivatives: in that case
 the structure of the
 higher-order terms in the low-energy action crucially depends on the quantum numbers of the $\Phi$'s.
 The leading two-derivative term defines relation (\ref{usual}) for the Goldstone decay constant as well as the leading self-interactions~\footnote{When the coset generators $T^A$ transform as a reducible representation of $\h$, in principle there will be a different $f$ for each quadratic invariant.}
  \bea
\frac{m_\rho^4}{g_\rho^2}{\cal L}^{0}&\equiv& f^2 {\rm Tr}\left( {\cal D}_\mu {\cal D}^\mu  \right)+\cdots\nonumber 
\label{quadratic}\\
&=&{f^2} {\rm Tr}\left [ \partial_\mu\Pi\partial^\mu \Pi +\frac{1}{3}
(\Pi {\overleftrightarrow{\partial_\mu}} \Pi)(\Pi {\overleftrightarrow{\partial^\mu}} \Pi)+\cdots\right ] .
\eea
Here ${\cal D}_\mu$ is the Goldstone combination defined in~\eq{DE} of appendix~A and $\Pi {\overleftrightarrow{\partial_\mu}} \Pi \equiv \Pi (\partial_\mu \Pi )-(\partial_\mu \Pi)\Pi$.
Once we interpret $\Pi$ as the Higgs doublet and include gauge covariant derivatives, we obtain that~\eq{quadratic} describes the following leading (dimension-6) interactions
\beq
 \frac{c_H}{2f^2}\partial^\mu \left( H^\dagger H \right) \partial_\mu \left( H^\dagger H \right) 
+ \frac{c_T}{2f^2}\left (H^\dagger {\overleftrightarrow { D^\mu}} H \right)  \left(   H^\dagger{\overleftrightarrow {D_\mu}} H\right) .
\label{dueoperatori} \eeq
Here we have  made a Higgs field redefinition
$H^\alpha \to H^\alpha +a (H^\dagger H)H^\alpha/f^2$ (with $a$ an appropriate constant) to write the operator 
$H^\dagger H |D_\mu H|^2$ in terms of the two appearing in~\eq{dueoperatori}. The coefficients
$c_H$ and $c_T$ are fixed by the $\sigma$-model structure, up to the overall normalization
which depends on the definition of $f$. For $SO(5)/SO(4)$ one finds $c_T/c_H=0$, because of custodial symmetry, and for
$SU(3)/SU(2)\times U(1)$ one finds $c_T/c_H=1$.

From eqs. (\ref{loopexpansion})--(\ref{quadratic}) we can deduce the rules to estimate the coefficients of the higher-dimensional operators in the low-energy effective Lagrangian
\begin{enumerate}
\item Each extra Goldstone leg is weighted by a factor $1/f$. For instance the addition of two Higgs doublet legs involves the factor
$H^\dagger H/f^2$.
\item Each extra derivative is weighted by a factor $1/m_\rho$. When the SM subgroup is weakly gauged, the replacement $\partial_\mu\to\partial_\mu+i A_\mu\equiv D_\mu$ is in order; this same rule implies that each extra insertion of
a gauge field strength $F_{\mu\nu}=-i[D_\mu,D_\nu]$ is weighted by a factor $1/m_\rho^2$.
\end{enumerate}

The global symmetry $\G$ is broken at tree level by the weak gauging of the SM group and by
the weak interactions that underlie the origin of Yukawa terms and Higgs potential.
 In sect.~\ref{secmod}, we shall present a more detailed analysis of all the various possibilities.  For our present goal we just need to remark
that if no new scale other than $m_\rho$ is present, 
  and simple expressions of the Goldstone field are involved,
  one expects  these breaking terms to satisfy the same field 
and derivative expansions expressed by rules 1 and 2. Basically, the selection rules of
$\G$ and of the flavour symmetry of the SM control the overall size of the symmetry breaking terms, while rules 1 and 2 determine the counting
for Higgs field and derivative insertions.  We can thus formulate rule 3:
\begin{enumerate}
\setcounter{enumi}{2}
\item  Higher-dimensional operators that violate the symmetry of the $\sigma$-model  must be suppressed by the same (weak) coupling associated to the corresponding renormalizable interaction in the SM Lagrangian ({\it e.g.}, Yukawa couplings $y_f$ and quartic Higgs coupling~$\lambda$).
\end{enumerate}
For instance,  the shift $H^\alpha \to H^\alpha +a (H^\dagger H)H^\alpha/f^2$ discussed before
induces the operators 
\beq
\left( \frac{c_yy_f}{f^2}H^\dagger H  {\bar f}_L Hf_R +{\rm h.c.}\right)
- \frac{c_6\lambda}{f^2}\left( H^\dagger H \right)^3 .
\eeq
The pure $\sigma$-model contributions give $c_y/c_H=-1/3$ and $c_6/c_H=4/3$ both in the cases of $SO(5)/SO(4)$ and $SU(3)/SU(2)\times U(1)$, with the definitions $y_f=\sqrt{2} m_f/v$ and $\lambda =m_H^2/(2v^2)$, valid up to corrections of order $v^2/f^2$.

 Of special phenomenological relevance  are the operators involving Higgs bosons and gauge
 fields, and in particular those involving one or two Higgses and a pair of photons or gluons like $O_{BB}=H^\dagger H B_{\mu\nu}B^{\mu\nu}$ and $O_{g}=H^\dagger H G_{\mu\nu}G^{\mu\nu}$.
 Using the CCWZ construction~\cite{ccwz}, we derive, in appendix~A, the following general structure of dimension-6 operators involving Higgs
and gauge field strengths, arising by weak gauging of the SM group:
\bea
 \label{OW}
&&O_W=i\left( H^\dagger  \sigma^i \overleftrightarrow {D^\mu} H \right )( D^\nu  W_{\mu \nu})^i
\qquad
 O_B=i\left( H^\dagger  \overleftrightarrow {D^\mu} H \right )( \partial^\nu  B_{\mu \nu}) \\\label{OHW}
 &&O_{HW}=i(D^\mu H)^\dagger \sigma^i(D^\nu H)W_{\mu \nu}^i\qquad
O_{HB}=i(D^\mu H)^\dagger (D^\nu H)B_{\mu \nu}\,.
 \eea
 While, expectedly,  operators involving gluons do not arise, it is also manifest that none of these operators contributes to the process $h\to \gamma\gamma$ (with real photons). On the other hand,
 there are contributions to $h\to Z \gamma$ from $O_{HW}$ and $O_{HB}$.
 Notice that, using  integration by parts, we could equivalently parametrize the set of operators in
 eqs.~(\ref{OW})--(\ref{OHW}) by
 $O_W$, $O_B$ and  $O_{ZB}=H^\dagger B^{\mu\nu}(W_{\mu\nu}+B_{\mu\nu})H $
 and $O_{ZW}=H^\dagger W^{\mu\nu}  (W_{\mu\nu}+B_{\mu\nu})H$, where $W_{\mu\nu}\equiv W^i_{\mu\nu}\sigma^i$. Again, neither $O_{ZB}$ nor $O_{ZW}$ contribute to $h\to \gamma \gamma$.
 According to
 counting rule 2, all these operators have a coefficient of order $1/m_\rho^2$, as they formally involve
 two extra covariant derivatives  with respect to a Higgs kinetic term. The absence of   $O_{g}=H^\dagger H G_{\mu\nu}G^{\mu\nu}$ and of operators affecting the
 coupling between Higgs and photons like  $O_{BB}=H^\dagger H B_{\mu\nu}B^{\mu\nu}$ is due to the Goldstone symmetry. 
Since the neutral Higgs $h$ is  both charge and color neutral, the  gauging of
 just $SU(3)_c\times U(1)_Q$  does not break the  $U(1)$ generator $T_h$ of $\G$ under which the physical Higgs boson shifts.
  Operators like $O_{g}$  or those leading to $h\to \gamma \gamma$ (like  $O_{BB}$) explicitly break this shift symmetry and cannot  be generated upon the simple  gauging of the SM group described by rule 2. 
In order  to generate these  terms, the couplings that break 
 the symmetry generated by $T_h$
must intervene, so that their coefficient must be suppressed by extra powers of $(\gsm/g_\rho)$.
Normally one gets a $\gsm^2/g_\rho^2$ extra suppression~\footnote{ A similar result holds in low-energy QCD. The coupling $\pi_0^2 
F_{\mu\nu}^2$ vanishes at leading order in $\alpha_{EM}$, and gets generated at subleading  order  in the quark mass $m_q$ and also through the chiral anomaly. On the other hand, $\pi^+\pi^- F_{\mu\nu}^2$
exists at zeroth order in both $\alpha_{EM}$ and $m_q$.}.

According to the general expression in
eq.~(\ref{loopexpansion}), four-derivative operators like those in eqs.~(\ref{OW})--(\ref{OHW})
can arise at tree level. However in ``normal'' theories, the classical action including the heavy fields $\Phi$ involves
at most two derivatives. Holographic Goldstone models and Little Higgs are of this type. To be more specific, these theories correspond to minimally-coupled field theories where the states have spin $\leq 1$,
and all vectors are associated to (spontaneously-broken) gauge symmetries~\footnote{The latter property sets the rule to count derivatives for massive vector fields through the requirement that the action for the eaten Goldstones be a 2-derivative one. For instance the gauge symmetry breaking term  $(\partial^\mu V_\mu)^2$ counts like a four derivative object and is discarded from the classical action. Minimal coupling along with the gauge principle ensures the absence of ghosts at the scale $m_\rho$ and a milder growth of the amplitudes at energies  above the scale $m_\rho$.}.
 In the case of minimally-coupled theories, higher-derivative operators
like those in eqs.~(\ref{OW})--(\ref{OHW}) can appear in the classical low-energy action
 below $m_\rho$ only if there exists a field $\Phi$ with the appropriate quantum numbers to mediate the corresponding operator. In this respect we remark an interesting difference between
 $O_W,O_B$ and $O_{HW},O_{HB}$. Two linearly independent  combinations of  $O_{HW}$ and $O_{HB}$ contribute  respectively to a vertex  that couples   an on-shell photon to two neutral states (a Higgs and a $Z$) and to a correction to the gyromagnetic ratio of the $W$.
 On the other hand, in minimally-coupled theories   photons do not interact 
 at tree level with neutral states  and all gyromagnetic ratios are fixed to be equal to $2$.  
 In these theories $O_{HW},O_{HB}$ must therefore appear in ${\cal L}^{(1)}$ and bear an extra one-loop suppression. By the same argument $O_{g}$ and $O_{BB}$  should also arise at one-loop and moreover, because of the previous symmetry-based argument, they should be further suppressed by a Goldstone symmetry-breaking power of $\gsm/g_\rho$. The operators $O_W$ or $O_B$
 can instead be generated in minimally-coupled theories by the tree-level exchange of heavy vector fields. We show this explicitly in appendix~A,
 in the context of a simplified model with a heavy vector that resembles both
 the Little Higgs models with product of gauge  groups and the Holographic Goldstone.
Given that all the known examples with $g_\rho< 4 \pi$ are minimally coupled, in the following of this paper, we will work under the assumption of minimal coupling~\footnote{One may wonder  how 
our results would change in a genuinely higher-spin (higher-derivative) theory like string theory. To be specific, one could consider, provided it exists, a realization of the Holographic Goldstone in weakly-coupled string theory  and take the limit $M_s\sim m_{KK}\equiv m_\rho$.
At first glace we would expect a drastic change. For instance, it is obvious that a photon can scatter off a dilaton, a neutral scalar, at tree level. On the other hand,  a specific study of  the gyromagnetic ratio $g$~\cite{Ferrara:1992yc} of all the high-spin states of the open string remarkably gives the result $g=2$, indicating a close similarity to a minimally-coupled theory.}. Of course this makes a difference when 
we consider models at $g_\rho< 4 \pi$, as opposed to the genuinely strongly coupled case $g_\rho\sim 4 \pi$, for which all loops are equally important in the spirit of NDA.

At the dimension-6 level, there is one last independent operator involving two Higgses and four covariant derivatives
\beq
\frac{1}{m_\rho^2}(D^2H^\dagger )(D^2 H) .
\eeq
As we show in the appendix~A this can be generated at tree level in a minimally coupled theory by integrating out a massive scalar transforming as an $SU(2)_L$ doublet. By the equations of motion
 this term can, however, be rewritten as 
\beq
\frac{1}{m_\rho^2}\left[ m_H^2H^\alpha+\lambda H^\dagger HH^\alpha +y_f(F_Lf_R)^\alpha \right]^2\, ,
\eeq
corresponding to effects 
that are all subleading to more direct corrections from the strong sector.

For completenes we should also list the dimension-6 operators involving only 
covariant derivatives and field strengths
\begin{eqnarray}
O_{2W}&=&(D^\mu W_{\mu\nu})^i (D_\rho W^{\rho\nu})^i 
~~~ O_{2B}=(\partial^\mu B_{\mu\nu})(\partial_\rho B^{\rho\nu})~~~ O_{2g} = (D^\mu G_{\mu \nu})^a  (D_\rho G^{\rho \nu )^a}\label{2v} \\
O_{3W}&=& \epsilon_{ijk} {W^{i}_\mu}^\nu W^{j}_{\nu\rho} W^{k\, \rho \mu}\qquad\qquad
O_{3g}= f_{abc} {G^{a}_\mu}^\nu  G^{b}_{\nu\rho} G^{c\, \rho \mu}.
\label{3v}
\end{eqnarray}
As we show in the appendix~A, see eq.~(\ref{intvectorI}), the three operators in \eq{2v} can be generated at tree level through the exchange of massive vectors transforming respectively as a weak triplet, as a singlet and as a color octet. Their coefficients are therefore in general of order $1/(g_\rho m_\rho)^2$. The two operators in \eq{3v}  cannot arise at tree level in
minimally-coupled theories. For instance $O_{3W}$ contributes to the
magnetic dipole and to the electric quadrupole of the $W$. They are thus generally expected with a coefficient $\sim 1/(4\pi m_\rho)^2$.

\subsection{The SILH effective Lagrangian}

We now basically have all the ingredients to write down the low-energy dimension-6 effective Lagrangian.
We will work under the assumption of a minimally coupled classical Lagrangian at the scale $m_\rho$.

Using the rules described in sect.~\ref{secrules}, we obtain a low-energy effective action for the leading dimension-6 operators involving the Higgs field of the form
\bea
&&{\cal L}_{\rm SILH} = \frac{c_H}{2f^2}\partial^\mu \left( H^\dagger H \right) \partial_\mu \left( H^\dagger H \right) 
+ \frac{c_T}{2f^2}\left (H^\dagger {\overleftrightarrow { D^\mu}} H \right)  \left(   H^\dagger{\overleftrightarrow D}_\mu H\right) \nonumber \\ &&
- \frac{c_6\lambda}{f^2}\left( H^\dagger H \right)^3 
+ \left( \frac{c_yy_f}{f^2}H^\dagger H  {\bar f}_L Hf_R +{\rm h.c.}\right) \nonumber \\ &&
+\frac{ic_Wg}{2m_\rho^2}\left( H^\dagger  \sigma^i \overleftrightarrow {D^\mu} H \right )( D^\nu  W_{\mu \nu})^i
+\frac{ic_Bg'}{2m_\rho^2}\left( H^\dagger  \overleftrightarrow {D^\mu} H \right )( \partial^\nu  B_{\mu \nu})  \nonumber \\
&&
+\frac{ic_{HW} g}{16\pi^2f^2}
(D^\mu H)^\dagger \sigma^i(D^\nu H)W_{\mu \nu}^i
+\frac{ic_{HB}g^\prime}{16\pi^2f^2}
(D^\mu H)^\dagger (D^\nu H)B_{\mu \nu}
 \nonumber \\
&&+\frac{c_\gamma {g'}^2}{16\pi^2f^2}\frac{g^2}{g_\rho^2}H^\dagger H B_{\mu\nu}B^{\mu\nu}+\frac{c_g g_S^2}{16\pi^2f^2}\frac{y_t^2}{g_\rho^2}H^\dagger H G_{\mu\nu}^a G^{a\mu\nu}.
\label{lsilh}
\eea
We will later discuss the Lagrangian terms that purely involve the vector bosons.
 The coupling constants $c_i$ are pure numbers of order unity. For phenomenological applications, we have switched to a notation in which gauge fields are canonically normalized, and gauge couplings explicitly appear in covariant derivatives. Also,
we recall the definition $H^\dagger {\overleftrightarrow D}_\mu H \equiv H^\dagger  D_\mu H- (D_\mu H^\dagger )H$.

In what follows, we will comment on the operators in~\eq{lsilh}. Let us start with the operators involving more than two Higgs fields. As previously discussed,
by using the Fierz identities for the Pauli matrices, one can write three independent operators involving four $H$ fields and two covariant derivatives. Two are shown in our Lagrangian with coefficients $c_H$ and $c_T$. The third operator
$H^\dagger H |D_\mu H|^2$, can be written in terms of a combination of $c_H,c_T,c_6,c_y$ by a Higgs field redefinition
$H^\alpha \to H^\alpha +(H^\dagger H)H^\alpha/f^2$, or, which is equivalent, by using the leading order equations of motion. 
The operator with coefficient $c_H$, as we will show in sect.~\ref{secphen}, plays a crucial role in testing the SILH in Higgs and vector boson scattering at high-energy colliders. The operator proportional to $c_T$ violates custodial symmetry and gives a contribution $\widehat T$ to the $\rho$ parameter
\beq
\Delta \rho\equiv \widehat T=c_T\xi,
\label{deltr}
\eeq
\beq
\xi \equiv \frac{v^2}{f^2}, \; \: \; v=\left( \sqrt{2} G_F\right)^{-1/2}=246\GeV .
\label{xi}
\eeq
From the SM fit of electroweak data~\cite{Barbieri:2004qk}, we find $-1.1\times 10^{-3}<c_T\xi<1.3 \times 10^{-3}$ at 95\% CL  (letting also $\widehat S$ to vary one finds instead $-1.7\times 10^{-3}<c_T\xi<1.9 \times 10^{-3}$ at 95\% CL). Because of this strong limit, we will neglect new effects from this operator and set $c_T$ to zero. Indeed, the bound on $c_T$ suggests that new physics relevant for electroweak breaking must be approximately custodial-invariant.
In our Goldstone Higgs scenario   this corresponds to assuming the coset $SO(5)/SO(4)$.  When $\gsm$ is turned on, $c_T$ receives a model dependent contribution, which should be small enough to make the model acceptable. In the next section, we will briefly discuss the
 size of $c_T$ in various models.

The coefficient $c_y$ is universal at leading order in the Yukawa couplings, and non-universal effects will appear at order $y_f^2/g_\rho^2$. This is because this term
purely originates from the non linearity in $H$ of the $\sigma$-model matrices. Indeed, the field redefinition mentioned
above precisely generates this universal $c_y$.

The operators proportional to $c_W$ and $c_B$ are generated respectively by tree-level exchange of a massive
triplet and singlet vector field as explained in the previous section (see also eq.~(\ref{intvectorI}) in appendix~A). Their relative importance in 2-to-2 scattering amplitudes with respect to the operator proportional to $c_H$ is $(g^2/g_\rho^2)(c_{W,B}/c_H)$. Therefore, in weakly-coupled theories ($g_\rho \sim g$), the two contributions are comparable but, in strongly-coupled theories
($g_\rho \gg g$), the operators proportional to $c_{W,B}$ give only subleading effects. Since, as we will show in sect.~\ref{secmod}, realistic models of electroweak breaking without excessive fine tuning prefer $g_\rho > g $, in most cases the contribution from $c_{W,B}$ are subleading with respect to the one from~$c_H$.

A linear combination of the operators with coefficients $c_W$ and $c_B$ contributes to the 
$\widehat S$~parameter of electroweak precision data:
\beq
	\label{eq:hatS}
\widehat S=\left( c_W+c_B\right) \frac{m_W^2}{m_\rho^2} ,
\eeq
where $\widehat S$ is defined in ref.~\cite{Barbieri:2004qk}.
Using the SM fit of electroweak data~\cite{Barbieri:2004qk}, we obtain the bound $m_\rho \gsim (c_W+c_B)^{1/2}~2.5$~TeV at 95\% CL.  (this bound
 corresponds to assuming a light Higgs and $\Delta \rho\equiv \widehat T=0$; by relaxing this request the bound becomes $m_\rho \gsim (c_W+c_B)^{1/2}~1.6$~TeV). In terms of the parameter $\xi$ defined in~\eq{xi}, this bound becomes
 \beq
 \xi \lsim\frac{1.5}{c_W+c_B} \left( \frac{g_\rho}{4\pi}\right)^2.
 \label{limxis}
 \eeq
 As we show in sect.~\ref{secphen}, new effects in Higgs physics at the LHC appear only for sizable values of $\xi$. Then~\eq{limxis} requires a rather large value of $g_\rho$, unless $c_W+c_B$ happens to be accidentally small.

The operators  with coefficients $c_{HW}$ and $c_{HB}$ originate from the 1-loop
action ${\cal L}^{(1)}$, under our assumption of minimal coupling for the classical action.
Although they are $H^2 D^4$ terms, like $c_{W},c_{B}$, they cannot be enhanced
above their 1-loop size by the exchange of any spin 0 or 1 massive field. In the case of a large $N$
theory where $N\sim 16\pi^2/g_{\rho}^2$, these terms are down with respect to the others by $1/N$.
Notice that according to this counting $\widehat S\sim g^2 N \xi /(16\pi^2)$, which for $\xi\sim 1$ coincides with the usual technicolor result. Recently, it has been pointed out that walking at small $N$ might be a promissing direction~\cite{Dietrich:2005jn}.

As discussed in sect.~\ref{secrules}, the operators proportional to $c_\gamma$ and $c_g$ are suppressed by an extra power  $(\gsm/g_\rho)^p$  with respect to those proportional to $c_{HW}$ and $c_{HB}$. Moreover, while $c_H$ and $c_y$ indirectly correct the physical Higgs coupling to gluons and quarks by $O(v^2/f^2)$ with respect to the SM, the direct contribution of $c_\gamma$ and $c_g$ is of order $(v^2/f^2)(\gsm/g_\rho)^p$. Their effect is then important only in the weakly coupled limit $g_\rho\sim \gsm$.
Notice that from the point of view of the Goldstone symmetry, $O_{BB}$ and $O_{g}$ are  like a Higgs mass term with extra field strength insertions. According to our power counting rules we then expect their coefficient to roughly scale like
${m_H^2}/{m_\rho}^4$ times the trivial factors of $g'^2$ and $g_3^2$. In the simplest models $m_H^2\sim (\gsm^2/16 \pi^2) m_\rho^2$. We have here assumed this simplest possibility, which accounts for the extra $\gsm^2/g_\rho^2$ appearing in \eq{lsilh}. More precisely,
for phenomenological purposes, we have chosen $\gsm $ as the coupling of the largest contribution in the corresponding SM loop, {\it i.e.}, $\gsm^2=g^2\ (y_t^2)$ for the operator involving photons (gluons), respectively.

Finally, for completeness, we give the dimension-6 Lagrangian  for  the vectors:
\begin{eqnarray}
{\cal L}_{vect}&=
&-\frac{c_{2W}g^2}{2g_\rho^2m_\rho^2}(D^\mu W_{\mu\nu})^i (D_\rho W^{\rho\nu})^i 
-\frac{c_{2B}{g'}^2}{2g_\rho^2m_\rho^2}(\partial^\mu B_{\mu\nu})(\partial_\rho B^{\rho\nu})-\frac{c_{2g}g_3^2}{2g_\rho^2m_\rho^2}(D^\mu G_{\mu \nu})^a  (D_\rho G^{ \rho \nu})^a\nonumber \\
&&+\frac{c_{3W}g^3}{16\pi^2 m_\rho^2} \epsilon_{ijk} {W^{i}_\mu}^\nu W^{j}_{\nu\rho} W^{k\, \rho \mu}+\frac{c_{3g}g_3^3}{16\pi^2 m_\rho^2}f_{abc} {G^{a}_\mu}^\nu  G^{b}_{\nu\rho} G^{c\, \rho \mu}\, ,
\label{vectors}
\end{eqnarray}
where the coefficients are dictated by the arguments given in sect.~\ref{secrules}.
 The operators proportional to $c_{2W}$, $c_{2B}$ and $c_{2g}$ arise from virtual tree-level exchange of massive vectors,
  from the second term in eq.~(\ref{intvectorI}) of appendix~A. The coefficients $c_{2W}$ and $c_{2B}$ contribute  to the electroweak precision
  parameters $W$ and $Y$~\cite{Barbieri:2004qk}
  \beq
  W= c_{2W} \frac{g^2m_W^2}{g_\rho^2m_\rho^2}\ ,\qquad\qquad Y= c_{2B} \frac{{g'}^2m_W^2}{g_\rho^2m_\rho^2}\,  .
\label{wwyy}
  \eeq
  By using the equations of motion for the SM field strengths, these operators can be rewritten
  as contact interaction among the electroweak currents.
  Given that the experimental bound on $W$ and $Y$ are
  comparable to that on $\widehat S$, in the moderately strong coupling regime $g_\rho > g$, the constaints on $c_{2W}$ and $c_{2B}$ are weaker than those on $c_W+c_B$. 

The Lagrangian terms in eqs.~(\ref{lsilh}) and (\ref{vectors}) include
14 dimension-6  CP-invariant operators involving only Higgs and gauge fields plus a 15th operator involving Higgs and  fermions, the one associated to $c_y$. This result, albeit in a different basis, agrees with ref.~\cite{Buchmuller:1985jz}
for the same class of operators.

\section{Relating the SILH to explicit models}
\label{secmod}

In this section, we consider a few explicit models that reduce 
to the effective Lagrangian~(\ref{lsilh}) below the masses of the new states. 
Before reviewing these models we would like to give a synthetic but comprehensive characterization of
the terms that explicitly break the Goldstone symmetry $\G$. We have already discussed in some detail the 
explicit breaking of $\G$ induced by the weak gauging of the SM group. 
Here  we will 
 discuss  the various possibilities for the Yukawa and Higgs couplings. 
 We can broadly distinguish
two classes of models: in the first class, the Higgs potential is fully saturated by quantum effects at the scale $m_\rho$, while in the second the quartic coupling is a tree-level effect (or, equivalently, it arises from quantum corrections at energies higher than $m_\rho$). 
The Georgi--Kaplan model and Holographic Goldstones are in the first class, while Little Higgses belong to  the second. 
In what follows we shall work with canonically normalized fields, and 
 indicate by $f_{L,R}$ the SM fermions while  by
$\Psi_{L,R}$ we refer to strong-sector states with mass $\sim m_\rho$.

\vskip0.5truecm
\noindent{\bf Class 1.}\
In this class of models, the full Higgs potential arises  
by  one-loop effects involving SM particles.
The dominant contribution is given by the  top quark   and its explicit form   
 depends  on the origin of the top Yukawa coupling.
We can  basically distinguish two possibilities to generate the  Yukawa couplings.
 The simplest one corresponds to models with minimal flavor violation~\cite{D'Ambrosio:2002ex} in which the only source of breaking of the $SU(3)^5$ global flavor symmetry of the SM are the Yukawa matrices
themselves. This case is realized  when the fermions couple to the strong sector bilinearly
\beq 
y_f\bar f_Lf_R O_S\, ,
\eeq
where $O_S$ is some operator of the strong sector, while 
$y_f$ are the SM Yukawa matrices. In the low-energy effective theory, the above term will give rise to the interaction 
\beq
y_f \bar f_L f_R H\,  P_y(H/f)\, ,
\label{inter}
\eeq
where $P_y$ is a polynomial whose expansion at $\mathcal{O}(H^2)$ determines the
universal coefficient $c_y$ of the effective Lagrangian~(\ref{lsilh}).
 Violations of $c_y$ universality will come  only at higher order in $y_f$.
By simple power counting, the
top-quark  contribution to the Higgs potential has the form 
 \beq
V(H)\sim \frac{m_\rho^4}{g_\rho^2}\times \frac{y_t^2}{16\pi^2}\times \hat V(H/f)\, .
\label{counting}
\eeq
Then the generic prediction of this class of models is $\langle H\rangle \sim f$ ({\it i.e.}, $\xi \sim 1$)  while
the Higgs quartic is $\lambda\sim y_t^2(g_\rho^2/16\pi^2)$. An upper bound around $m_t$ is thus
predicted for the Higgs mass, in some analogy with  supersymmetry. 
Notice that,  in this type of models, it is mandatory  that  $g_\rho\gg \gsm$ 
in order to have a Higgs mass above the experimental bound.
Notice also that
\beq
\widehat S\sim \frac{\gsm^2}{g_\rho^2}\xi\, ,
\eeq
which  for maximal strength $g_\rho\sim 4\pi$ and $\xi \sim 1$ has the size of an electroweak loop. 
(By interpreting $N \sim (4\pi/g_\rho)^2$, the generic prediction for $\widehat S$ qualitatively
corresponds to the case of $N$ technicolors.)
Then the smaller $g_\rho$, the more tuning on $\xi$ is needed in order to satify  the experimental bound on $\widehat S$. These models must therefore largely satisfy our assumption $g_\rho \gg \gsm$. 
On the other hand, for $g_\rho$ somewhat less than maximal we must tune $\xi$ by an amount $ (g_\rho/4\pi)^2<1$.  In order to achieve that, we clearly need extra contributions to the potential, associated to other 
$\G$ breaking couplings, possibly involving only the heavy states.
Provided the extra contributions have different form than the one from the top quark, we may tune a little bit the quadratic term with respect to the quartic, thus suppressing $\xi$.

The second possibility to generate  Yukawa couplings  is to have the   SM fermions couple linearly to fermionic  operators of the strong sector:
\beq
y_L \bar f_L O_R+y_R \bar f_R O_L +{\rm h.c.}\, ,
\label{inte}
\eeq
where  $y_{L,R}$ are matrices in flavor space. 
In the simplest cases, $O_{L,R}$ have definite quantum numbers under $\G$, and
therefore~\eq{inte} formally determines the spurionic quantum numbers   of $y_{L,R}$.
The possibility of generating Yukawas from  the linear couplings of~\eq{inte} 
  was first   suggested  in ref.~\cite{Kaplan:1991dc}
 for Technicolor models,
and it is the one    implemented  in   Holographic Higgs models~\cite{Agashe:2004rs}.
Writing~\eq{inte}
as a function of the physical  states of the strong sector $\Psi$,    
one can see that in these models  the Yukawa couplings are  
generated  through a sort of universal see-saw
\beq
m_\rho\left [ \frac{ y_L}{g_\rho} \bar f_L \Psi_R P_L(H/f) +\frac{y_R}{g_\rho} \bar f_R \Psi_L P_R(H/f)+ \bar \Psi_L\Psi_R\right ] \, .
\label{lightheavy}
\eeq
Notice  that  for $y_{L}\sim g_\rho$ or $y_{R}\sim g_\rho$ respectively $f_L$ or $f_R$ should be considered  as part of the strong sector~\footnote{For instance, for $y_R\sim g_\rho$ the linear combination of
$f_R$ and $\Psi_R$ which is left massless by the second and  third terms in~\eq{lightheavy} has the natural interpretation of a massless composite.}. This remark explains
the normalization of the mixing term in~\eq{lightheavy}.
The effective SM Yukawa couplings after integrating out the $\Psi$ have  the form
\beq
y_f\sim \frac{y_L y_R}{ g_\rho}\, .
\eeq
For $y_L\sim y_R$ one has $y_{L,R}\sim\sqrt{ y_f g_\rho}$, which is a coupling of intermediate strength.
 If the polynomials $P_L$ and $P_R$ are flavor universal so will be the $c_y$ coefficient at leading order in the $y_{L,R}$. Nevertheless, 
 the exchange of $\Psi$   will give rise to non-universal ($H$ dependent) corrections to the kinetic terms of $f_{L,R}$ that will scale like
 $y_{L,R}^2/g_\rho^2$. 
 By going to canonically normalized fermions one induces then 
 $\mathcal{O}(y_{L,R}^2/g_\rho^2)$ non-universal corrections to $c_y$. 
The top contribution to the Higgs  potential receives now,
in addition to  terms of the form (\ref{counting}),  corrections  scaling like
\beq
V(H)\sim \frac{m_\rho^4}{g_\rho^2}\times \frac{y_{L,R}^2}{16\pi^2}\times \hat V(H/f)\, .
\label{potLR}
\eeq
For $y_L\sim y_R$ this leads to 
a Higgs quartic coupling  $\lambda \sim (g_\rho/4\pi)^3 4\pi y_t$, and therefore a  moderately heavy Higgs boson ($\sim 300 $ GeV) can in principle be obtained~\footnote{Whether this can be achieved in practice depends on the specific model at hand. Depending on the $\G$ quantum numbers of $y_{L,R}$, the potential they generate may or may not trigger electroweak symmetry breaking. For instance in the
model in ref.~\cite{Agashe:2004rs}, the Higgs potential terms \eq{potLR}  align to the wrong  vacuum  $\langle H\rangle =0$. In that model the formally subdominant genuine top contribution \eq{counting},
which has a minimum at $\langle H\rangle \sim f$, must therefore be equally important. This is easily achieved by some little tuning of $y_L$ and $y_R$ thanks to an accidental numerical suppression of \eq{potLR}. The result however is that the Higgs mass is bounded $\lsim m_t$.}. For the same reason the suppression of
the coefficients of $O_{g}$ would be $y_t/g_\rho$ instead of
$y_t^2/g_\rho^2$.
Obviously, in the particular case in which   the right-handed top is a 
singlet under  the global group $\G$, its 
 contribution to the Higgs potential~\eq{potLR} will vanish.
The allowed values for the couplings $y_{L,R}$ strongly depend
on    the  quantum numbers of the
mixing operators $O_{L,R}$.
In the simplest  case in which 
$O_L=(2,1)$, $O_R=(1,2)$ under the custodial group 
$SO(4)=SU(2)_L\times SU(2)_R$, 
the corrections to $Z\bar bb$ and $\widehat T$ are expected to be~\footnote{Notice 
that  $\widehat T$  corresponds to a  contribution to the vector boson mass matrix transforming as an object with custodial isospin $I_c= 2$. 
Therefore,   $y_{L}$,  being  an isospin singlet,
 will not contribute to $\widehat T$, 
while $y_R$, being  a doublet of $SU(2)_R$, will have to  
  enter at least at fourth order.}
\begin{equation}
\frac{\delta g_b}{g_b}\sim \frac{y_L^2}{g_\rho^2}\xi\ ,\ \ \ \ \ 
 \widehat T= 
  \frac{N_c y^4_{R}}{16\pi^2g_\rho^2}\xi\label{deltabdeltaT}\, ,
\end{equation}
where $N_c=3$ is the number of colors.
The experimental  bounds, together with  the relation $y_t\sim y_L y_R/g_\rho$   imply $\xi<0.05$. The reason of this tight bound is that  $\delta g_b/g_b$ demands a small $y_L$,
 $\widehat T$ demands a small $y_R$,   while the two couplings are constrained to have a sizeable product to reproduce $y_t$. A less constrained,
and thus less tuned scenario, can arise in the less minimal case where
$O_L=(2,2)$, $O_R=(1,1)$. Now $y_R$ is a singlet
under  the custodial group and  drops out of~\eq{deltabdeltaT}.
However $y_L$  transforms as $(1,2)$ under $SU(2)_L\times SU(2)_R$ and therefore 
\begin{equation}
\frac{\delta g_b}{g_b}\sim \frac{y_L^2}{g_\rho^2}\xi\ ,\ \ \ \ \ 
 \widehat T\sim  \frac{N_cy^4_{L}}{16\pi^2g_\rho^2}\xi\, .
 \label{deltabdeltaT11}
\end{equation}
For $y_L\sim y_t$, corresponding to a composite $t_R$ 
($y_R\sim g_\rho$),
the bound from $\delta g_b/g_b$ is  comparable to the one from $\widehat S$, while the one from $\widehat T$ is less severe. In this case, however, the mass of the Higgs is again bounded to be parametrically $\lsim m_t$. 
A larger $y_L$, and  a possibly heavier Higgs, can 
remarkably be made compatible with $\delta g_b/g_b$ for the special case
in which the theory possesses an additional parity $P_{LR}$ exchanging the $SU(2)_L$ and $SU(2)_R$ group within $SO(4)$. In that case, the leading tree-level contribution to $\delta g_b/g_b$ can be naturally set to zero, and one can take $y_L\sim\sqrt{y_t g_\rho}$
without  being in stark conflict with the bound on $\widehat T$. For this value of $y_L$ flavor violating effects
are of the order of  the experimental bounds, as we will see in  sect.~\ref{sectop}. Of course our previous warnings about obtaining proper electroweak breaking still apply here.

Apart from the above exceptional case in which $O_L=(2,2)$, $O_R=(1,1)$,
other  alternatives for the quantum numbers of  $O_{L,R}$
will generically have  problems  with $\delta g_b/g_b$ and~$\widehat T$.
We can
take  $y_L\sim y_t$ and $y_R\sim g_\rho$ 
in order to     reduce $\delta g_b/g_b$,
 but in this case  $\widehat T$ comes always too large, $\widehat T\sim N_c g^2_\rho\xi/(16\pi^2)$.
A possible  way to reduce $\widehat T$  is  to  introduce into the theory 
custodial partners for the top whose  masses  $m_{cust}$ will control the breaking 
of the custodial symmetry \cite{Contino:2006qr}.  
In this case  the contribution to $\widehat T$   will have  extra supression factors
$(m_{cust}/m_\rho)^2$   that can reduce $\widehat T$ below the experimental bound 
for  $m_{cust}\sim m_t$~\footnote{One must, however,  check that these extra states
do not have electromagnetic charge $Q=-1/3$ and 
mix with $b_L$, since this would induce  large effects on $\delta g_b/g_b$.
This can occur, for example,    for  the assignment  
$O_L=(3,2)$ and  $O_R=(1,2)$.}.
This possibility  implies the presence of extra light fermions  that are easily accessible at the LHC or even at Tevatron.

\vskip0.5truecm
\noindent{\bf Class 2.} These are models realizing the clever Little Higgs  construction by which only the quadratic term in the Higgs potential is saturated by  quantum corrections at the $m_\rho$ scale. 
The Higgs quartic  term, in particular, is sensitive to the larger  scale $\Lambda\sim 4\pi f$ and is estimated to be of order
 $\lambda \sim \gsm^2\Lambda^2/(16\pi^2 f^2)\sim \gsm^2$. 
We then have
\beq
V(H)\sim \frac{\gsm^2}{16\pi^2} m_\rho^2 H^2 +\gsm^2 H^4\, .
\label{LHpot}
\eeq
The minimization yelds the parametric relation  $v^2/f^2 \sim g_\rho^2/(16 \pi^2)$ to be compared to  $v^2/f^2 \sim 1 $ for  models of Class 1.
The relation $\lambda \sim \gsm^2$  implies that 
the Higgs mass can now be above the experimental bound without requiring a large $g_\rho$,  in contrast to the   models of Class 1.  In a related way we have roughly 
$\widehat S \sim {m_W^2}/{m_\rho^2}={g^2}/{16\pi^2}$,
which also does not manifestly push towards a large $g_\rho$. Now, the result for $\widehat S$ in the Little Higgs is parametrically the same of a 1-loop electroweak correction, which is typically only marginally acceptable. Better agreement with the bound on $\widehat S$ can be achieved by using the flexibility the Litlle Higgs possesses 
in the strong sector parameters:
$g_\rho$ truly represents a spectrum of different couplings. In particular we can distinguish a coupling $g_\rho$ associated to the extra gauge factors and a coupling $g_T$ associated with the mass $m_T\sim g_T f$ of the partners of the top quark. Since $\widehat S$ is controlled by the vector boson mass $m_\rho\sim g_\rho f$, while the scalar potential tends to be dominated by the top contribution, the expression for $\widehat S$ is more appropriately
\beq
\widehat S \sim \frac{m_W^2}{m_\rho^2}\sim \frac{g^2}{16\pi^2}\frac{g_T^2}{g_\rho^2}\, ,
\label{SLH}
\eeq
showing that there is space to relax $\widehat S$ with respect to its typical one-loop size,  making it  numerically acceptable~\footnote{Notice that if we take
$ g_\rho\gsim  g_T y_t/g$ the vector boson loops dominate the mass term in the potential, in which case either
the relaxation effect saturates or the mass  term changes sign thus restoring electroweak symmetry. Thus we can naturally relax $\widehat S$ down to
 $\sim (g^4/y_t^2)/16\pi^2$, which can  be numerically acceptable.}.
Therefore this class of models prefers  a  weak coupling  in the top resonance sector 
but still a somewhat large coupling $g_\rho$ in the new gauge sector. 

Concerning $\widehat T$,
if the underlying $\sigma$-model does not posses a custodial
symmetry, one will have a significant $\widehat T \sim \mathcal{O}(v^2/f^2)\sim g_T^2/16\pi^2$
requiring a severe tuning of parameters and disfavoring large $g_T$ even more. But even when the model is custodially symmetric in the limit $\gsm=0$, there are corrections to $\widehat T$ that are potentially important at small $g_\rho$.  This is because SM custodial breaking couplings like $y_t$
modify at tree level the structure of the strong sector Lagrangian, in particular the potential of electroweak triplet scalars (these effects unlike the Higgs mass are not screened above the scales $m_T$, $m_\rho$,  and like the Higgs quartic they formally have tree-level size). 
If the leading contribution to the mass of the triplets is controlled by $g_\rho$,
then, by  the isosping argument illustrated before, we generally expect $c_T\sim (y_t/g_\rho)^4$ in \eq{lsilh},
leading roughly to $\widehat T\sim y_t^4 g_T^2/(16\pi^2 g_\rho^4)$.  This also favors $g_\rho >g_T, y_t$. A more detailed scrutiny of these effect requires considering some explicit model, which is beyond the scope of our brief survey. 

Notice that the general class of $O(\gsm^2/g_\rho^2)$ (and also  $\gsm^2/g_T^2$) effects including those we have just discussed, would, for $\gsm\sim g_\rho$, induce $O(1)$ violations of the $\G/\h$ $\sigma$-model structure below the scale $m_\rho$. For instance, in  addition to $c_T$, the terms of dimension higher than 6 in the two derivative Higgs Lagrangian would not be those dictated by the $\G/\h$ structure. 
In realistic LH models a weak $g_T\sim y_t$ is somewhat favored. Then in the contribution of the top partners to the Higgs-gluon coupling $c_g$  the Goldstone suppression $y_t^2/g_T^2$
is $O(1)$, and this operator is as important as $c_H$ and $c_y$ in Higgs physics.  
 Notice however that in the weakly-coupled limit,
 all anomalous effects in Higgs physics are of order $v^2/f^2\sim (g_T/4\pi)^2$, which is  parametrically like a SM loop effects.
  Expectedly, the use of our effective Lagrangian is  more motivated,  the more strongly coupled the new sector is, that is the bigger $v/f$.

In the following,
we will give three explicit examples of models with the above characteristics.
We will first concentrate on the Holographic Composite Higgs model of~\cite{Agashe:2004rs,Contino:2006qr}.
Then we move to the Littlest Higgs model~\cite{Arkani-Hamed:2002qy} and finally to  a Little Higgs model with custodial symmetry~\cite{Chang:2003zn}.

\subsection{Holographic composite Higgs model}

The Holographic Higgs model~\cite{Agashe:2004rs,Contino:2006qr}
is based on a  five-dimensional  theory in  AdS space-time. 
This space-time, of  constant    radius of curvature $1/k$,   is  assumed to be 
compactified by two 4D boundaries.  
One boundary is located  at $z=L_0$ (where $z$ labels the extra dimension in conformal coordinates)
and it is referred as the  UV boundary, while the other one is    at $z=L_1\gg L_0$ and it is called
the IR boundary. The energy scale $1/L_1$ 
sets the mass gap of the model (the Kaluza--Klein mass $\sim 1/L_1\sim 1$ TeV).
The bulk gauge symmetry $SO(5)\times U(1)_X\times SU(3)$ is broken
down to the SM gauge group on the  UV boundary and to $O(4)\times U(1) _X\times SU(3)$
on the IR. 
The hypercharge  is defined by $Y=X+T_3^R$ where $X$ is the $U(1)_X$ charge and 
$T_3^R$  is the 3rd-component isospin of $SU(2)_R\in O(4)$.

We will follow  the Holographic approach and  separate the 5D  gauge fields  in 
UV-boundary fields $A(z=L_0,x)$ and bulk  fields $A(z\not=L_0,x)$. 
This is the correct separation  to make contact with the theory defined in  sect. 2;
the UV-boundary  fields   can be associated
to the SM gauge bosons, while  the bulk states   correspond to the 
 new ``strong" sector.
Let us analyze this new sector.
Since the symmetry-breaking pattern of the  bulk and IR-boundary
 is given by $SO(5)\rightarrow O(4)$,  we expect
four  Goldstone bosons   parametrized by  the $SO(5)/SO(4)$  coset~\cite{Agashe:2004rs}:
\begin{equation}
\Sigma = \langle \Sigma\rangle e^{\Pi/f} \ ,
 \qquad \langle \Sigma\rangle =  (0,0,0,0,1) \ ,
 \qquad 
 \Pi = 
\left(
\begin{array}{cc}
0_{4} & H \\
- H^T & 0\\
\end{array}
\right)
\, ,
\end{equation}
where $H$ is  a  real 4-component vector, which transforms as a doublet 
under the weak $SU(2)$ group and can be associated with  the Higgs.
Apart from the Goldstones, the bulk  contains a massive tower of 4D states, 
 the  gauge Kaluza--Klein  modes.
The mass of the lightest state  is given by 
\begin{equation}
m_\rho\simeq \frac{3\pi}{4L_1}\ ,\ \ \ \ {\rm and}\ \ \ 
g_{\rho}=\frac{m_\rho}{f}\simeq\frac{3\pi}{8}\sqrt{g^2_5 k}\, ,
\end{equation}
  where $g_5$  is the bulk $SO(5)$  gauge coupling.
In the particular case where  the  red-shift factor between the two boundaries $L_0/L_1$ 
is used to  explain the hierarchy $m_W/M_P$, 
we obtain that $g_5^2 k\gsim g^2\ln(L_1/L_0)\sim 16$
implying that  the coupling among resonances is always large.
As we explained before, a large $g_\rho$ is also needed in these models to 
guarantee a    Higgs mass  above the experimental bound.  
The interaction between 
the massive Kaluza--Klein states  and the fields on the UV-boundary (the SM fields)
  is due only to mass mixing  terms.
  These terms  only  respect the SM gauge symmetry and therefore corresponds to an explicit
breaking of the $SO(5)$ symmetry. 
For the fermion sector, we will follow ref.~\cite{Contino:2006qr}  and  embed the SM 
fermions in the ${\bf 5}$ representation of $SO(5)$.  
We can again separate each  5D fermion  in a UV-boundary field, 
to be associated to the SM fermion, plus a bulk field.
As in \eq{lightheavy}, Yukawa couplings are generated in this model 
through  mass mixing terms  between   the  SM fermions 
and   the heavy fermionic bulk modes.
The size of these mixing couplings   are determined by the 
5D fermion masses  and  can be chosen
to give  the correct SM   spectrum.

Let us calculate the contribution  of this model to  
the coefficients of the effective operators of eq.~(\ref{lsilh}). 
The  coefficients   $c_T$ and $c_H$ can be obtained from  the kinetic term of  the Goldstone bosons:
\begin{equation}
{\cal L}_{\rm kin}=\frac{f^2}{2}(D_\mu\Sigma)(D^\mu\Sigma)^T\, .
\label{sigma5}
\end{equation}
In  the unitary gauge where
$
\Sigma = \left(\sin h/f ,0,0,\cos  h/f \right)
$, \eq{sigma5} gives
\beq
{\cal L}_{\rm kin}=\frac{1}{2}\partial_\mu h\partial^\mu h+m^2_W(h) \left[ W_\mu W^\mu+\frac{1}{2\cos^2\theta_W}
Z_\mu Z^\mu\right]\, , 
\label{sigma52}
\eeq
where
\beq
m_W(h)=\frac{gf}{2}\sin\frac{h}{f}\, .
\label{mwh}
\eeq
Eq.~(\ref{sigma52}) tells us that   $\Delta\rho=0$  or, equivalently, $c_T=0$.
This is due to the  custodial $O(3)$ invariance of \eq{sigma5}.
The value  of $c_H$  can be unambiguously computed by comparing the 
$hWW$ coupling for canonical fields in \eq{sigma52}  (neglecting $m_W^2/m_\rho^2$ corrections) with the same
quantity deduced from our general effective Lagrangian \eq{lsilh}.
This second step, which requires writing \eq{lsilh} in  the same field basis as \eq{sigma52}, is performed in Appendix B.
 From \eq{sigma52} we  have
\begin{equation}
\frac{g_{hWW}}{gm_W}=
\left.\frac{1}{gm_W(h)}\frac{\partial m^2_W(h)}{\partial h}\right|_{h=\langle h\rangle}=\cos\frac{\langle h\rangle}{f}\simeq
1-\frac{\xi}{2}\, ,
\label{hwwh}
\end{equation}
and,  using   \eq{hww},  we find
\begin{equation}
c_H=1\, .
\label{ch}
\end{equation}
The coefficient $c_y$ can be similarly deduced  from the calculation of the  $hff$ coupling.
For the  Holographic Higgs model of  ref.~\cite{Contino:2006qr}
we have 
\be
{\cal L}_{\rm yuk}=-m_f(h)\bar ff\ ,\qquad  m_f(h)= M \sin\frac{2h}{f}\, ,
\label{yuk}
\ee
where $M$ is a constant. 
We then obtain
\begin{equation}
\frac{2m_Wg_{hff}}{gm_f}=
\left.\frac{2 M_W(h)}{gm_f(h)}\frac{\partial m_f(h)}{\partial h}\right|_{h=\langle h\rangle}=
\frac{2\sin(\langle h\rangle/f)}{\tan(2\langle h\rangle/f)}\simeq 1-\frac{3\xi}{2}\, ,
\end{equation}
that, comparing  it with eq.~(\ref{hff}), leads to 
\begin{equation}
c_y=1\, .
\end{equation}
To obtain  the coefficient $c_6$ we must match the  $hhh$ coupling  obtained from the Higgs potential to \eq{hhh}.
In the model of ref.~\cite{Contino:2006qr} in which the Higgs potential can  approximately 
be written as $V(h)\simeq \sin^2h/f[\alpha-\beta\cos^2h/f]$ (where $\alpha$ and $\beta$
are constants), we have
\beq
\left.\frac{4m_Wg_{hhh}}{gm^2_H}=\frac{4m_W(h)}{g\partial^2_hV(h)}\frac{1}{6}\frac{\partial^3V(h)}{\partial h^3}\right|_{h=\langle h\rangle}=\frac{1-2\sin^2(\langle h\rangle/f)}{\cos(\langle h\rangle/f)}\simeq 1-\frac{3\xi}{2}\, .
\label{h3h}
\eeq
 From \eq{ch},   (\ref{h3h}) and (\ref{hhh})  we get
 \beq
 c_6=0\, .
 \eeq
In  the model of ref.~\cite{Agashe:2004rs}, in which   the SM fermions are embedded in
the spinorial representation  of $SO(5)$,  we find
 $c_y=0$ and   $c_6=1$.

The coefficients $c_{W,B}$, $c_{2W}$ and $c_{2B}$
can be obtained respectively from the parameters $\widehat S$, $W$ and $Y$. 
The parameter $\widehat S$, 
at tree level, is given by \cite{Agashe:2004rs}
\begin{equation}
\widehat S= \frac{3g^2\xi}{8g^2_5}\simeq \frac{27\pi^2}{128}\frac{m^2_W}{m_\rho^2}\, .
\label{shat}
\end{equation}
Using eq.~(\ref{eq:hatS}) and \eq{shat} together with the fact that the $O(4)$ symmetry of the model implies $c_W=c_B$, we obtain
\beq
c_W=c_B= \frac{27\pi^2}{256}\simeq 1.0\, .
\label{cw}
\eeq
For the  parameters $W$ and $Y$  we obtain
\beq
 W= \frac{g^2}{4g^2_5}m^2_WL_1^2\ ,\qquad
  Y=\frac{g^{\prime\, 2}}{4}\left(\frac{1}{g^2_5}+\frac{1}{g^2_{5X}}\right)m^2_WL_1^2\, ,
\label{xy}
\eeq
  where $g_{5X}$  is the bulk U(1)$_X$  gauge coupling.
Using  \eq{wwyy}, we get
\beq
c_{2W}\simeq \left(\frac{9\pi^2}{64}\right)^2\simeq 1.9\ ,\qquad   c_{2B}= c_{2W}\left(1+r^2\right) \, ,
\eeq
where $r=g_5/g_{5X}$.
The coefficients
  $c_{HW,HB}$ and  $c_{\gamma,g}$ 
  will not be presented here.
They are generated   at the  one-loop level  and are therefore   very sensitive to 
the details of the 5D model. 
Similarly, the non-universal contribution to $c_y$, although generated at tree-level,   
depends on the   particular structure  of the top sector
and will be discussed in sect.~\ref{sectop}.

Although the  calculations for
$c_{W,B}$ and  $c_{2W,2B}$ (and therefore $\widehat S$, $W$ and $Y$) 
are only valid for   $g_\rho< 4\pi$,
 the predictions  for $c_H$, $c_T$,  $c_y$  and $c_6$ at leading order
can be trusted even in the non-perturbative regime.
The coefficients  $c_H$ and  $c_T$ are, as we said,  completely determined 
by   the symmetry breaking pattern of the model,  
while  $c_y$  and $c_6$ depend 
only on the way we embed   the SM fermions into the SO(5) group.
These coefficients  are therefore   independent  of the five-dimensional dynamics.

\subsection{Littlest Higgs model}

The Littlest Higgs model~\cite{Arkani-Hamed:2002qy} is based on a global $SU(5)$ symmetry and we consider the version where only a $SU(2)_L\times SU(2)_R \times U(1)_Y$ subgroup of $SU(5)$ is gauged ($g_L$, $g_R$ and $g'$ are the respective gauge couplings). As we are going to show, below the scale of the heavy new particles, this model can be described by our effective Lagrangian with a $SU(3)/SU(2)$ coset structure.

It is assumed that a UV dynamics breaks the global $SU(5)$ symmetry down to $SO(5)$. This breaking is conveniently parametrized in terms of a $SU(5)$ symmetric representation acquiring a vev of the form
\begin{equation}
\langle \Sigma \rangle = 
\left(
\begin{array}{ccc}
& & \mathbf{1}_2\\
& 1\\
\mathbf{1}_2 \\
\end{array}
\right) .
\end{equation}
Among the 14 Goldstone bosons, 3 are eaten in the breaking of the gauge symmetry $SU(2)_L \times SU(2)_R$ down to the diagonal subgroup $SU(2)$ and we are left with a charged doublet $H_{1/2}$, a charged triplet $\phi_{1}$ and a neutral singlet $s_0$ (the subscripts denote the $U(1)_Y$ charges of the fields). These Goldstones are parametrized by
\begin{equation}
	\label{eq:PGBSU5}
\Sigma= e^{i\Pi/f}\, \langle \Sigma \rangle\,  e^{i\Pi^T /f} 
= 
e^{2i\Pi/f} \langle \Sigma \rangle
\hspace{.5cm} \mathrm{with}\hspace{.5cm} 
2\sqrt{2} i\Pi = 
\left(
\begin{array}{ccc}
is & {\tilde H} & \phi\\
- {\tilde H}^\dagger & -4is & {\tilde H}^T\\
-\phi^\star & -{\tilde H}^\star & is \\
\end{array}
\right).
\end{equation}
$\phi$ is a $2\times 2$ symmetric complex matrix, ${\tilde H}=i\sigma^2 H^\star$ and  $f$ is the decay constant of the coset model. The interactions of the Goldstones originate from the kinetic term of the $\Sigma$ field
\begin{equation}
\mathcal{L} = \frac{f^2}{2} \tr D_\mu \Sigma^\dagger D^\mu \Sigma ,
\end{equation}
where the covariant derivative accommodates the gauging of $SU(2)_L\times SU(2)_R \times U(1)_Y$ 
\begin{equation}
D_\mu \Sigma = \partial_\mu \Sigma - i g_L (A^L_\mu \Sigma + \Sigma {A^L_\mu}^T)
- i g_R (A^R_\mu \Sigma + \Sigma {A^R_\mu}^T)
- i g' B_\mu (Y \Sigma + \Sigma Y) ,
\end{equation}
with
\begin{eqnarray}
& \displaystyle
A^L_\mu = A^{a\, L}_\mu \left( \begin{array}{ccc} \sigma^a /2 & \phantom{0} & \phantom{0}\\ \\ \\ \end{array}\right)\ , 
\hspace{.3cm}
A^R_\mu = A^{a\, R}_\mu \left( \begin{array}{ccc}  \\ \\  \phantom{0} & \phantom{0} & - \sigma^{a\, \star}/2 \end{array}\right)\, ,\\[.3cm]
& \displaystyle
Y=\textrm{diag} \left(-\frac 12,-\frac 12 ,0,\frac 12 ,\frac 12 \right)\,  .
\end{eqnarray}

The $\langle \Sigma \rangle$ vev gives a mass to the axial part of $SU(2)_L\times SU(2)_R$,
$m_{W_H}^2 = (g_L^2+g_R^2) f^2$. While the gauge coupling of the unbroken vectorial $SU(2)$ takes its usual expression $1/g^2=1/g_L^2+1/g_R^2$. By the construction of the model, when the $SU(2)_L$ gauge coupling is turned off, the gauging of $SU(2)_R$ respects a $SU(3)$ global symmetry and  the doublet and the singlet are exact Goldstone bosons while the charged triplet acquires a mass of order $ g_R f$.  Below the scale $g_R f$, the Littlest Higgs model has the structure described in sect.~\ref{secstruc} of a $SU(3)/SU(2)$ $\sigma$-model weakly coupled to a $SU(2)_L\times U(1)_Y$ gauge sector.
 According to our description we should then identify
$g_\rho\equiv g_R$ and the mass of the axial vector is $m_\rho$.

After integrating out $A_R$, as it is explained formally in appendix~A, as well as the charged scalar triplet~\footnote{At the $1/f^2$ order, the integration of the scalar triplet is equivalent to the constraint $\phi^\star = \frac{{\tilde H}{\tilde H}^T}{\sqrt{2}f}$.},
the low-energy effective Lagrangian can be mapped onto the $SU(3)/SU(2)$ Lagrangian
\begin{equation}
2 f^2 \tr [\partial_\mu \Sigma_3^\dagger \partial^\mu \Sigma_3] + \frac{1}{2} f^2 \tr | \Sigma_3^\dagger \partial_\mu \Sigma_3|^2,
\end{equation}
where $\Sigma_3$ parametrizes the $SU(3)/SU(2)$ coset. This totally fixes the relative coefficient between the two independent invariants that exist in $SU(3)/SU(2)$ due to the fact that this coset decomposes into 2 irreducible representations of $SU(2)$.

The coefficients $c_i$ of the corresponding  SILH Lagrangian can be computed along the lines outlined in the previous subsection. 
The oblique parameters are found to be, in the limit $g_\rho \gg g$,
\begin{equation}
{\widehat S} = \frac{m_W^2}{2 m_\rho^2}\ ,\hspace{.2cm}
{\widehat T} = -\frac{m_W^2}{4 m_\rho^2}\ ,\hspace{.2cm}
W= \frac{g_L^2}{g_R^2} \frac{m_W^2}{m_\rho^2}\ ,\hspace{.2cm}
Y= \mathcal{O} (v^6/f^6)\, .
\end{equation}
From this and eqs.~(\ref{deltr}),  (\ref{eq:hatS}), (\ref{wwyy}) we deduce that
\begin{equation}
c_W=\frac 12\ ,\ \  c_B=0\ ,\  \  c_T=-\frac{1}{16}\ , \ \  c_{2W}=1 \  ,\ \   c_{2B}=0,
\end{equation}
where we have taken $c_B=0$, since the only gauge fields integrated out form an adjoint of $SU(2)$.
The value of $c_H$ can  be easily deduced from
the $hWW$ coupling at zero momentum:
\begin{equation}
g_{hWW}  = g m_W \left( 1 - \frac{\xi}{8} \right)\,,
\end{equation}
that,  together with   \eq{hww},  leads to
\begin{equation}
c_H=\frac{1}{4}\,  .
\end{equation}
Notice that the low-energy $\sigma$-model, $SU(3)/SU(2)$, breaks custodial symmetry  and hence we obtain a non-vanishing $c_T$ coefficient. The source of this custodial breaking is the vev of the triplet $\phi$. It can always be fine-tuned away, for instance by taking $g_R \sim g_L$. In that case
\begin{equation}
c_T=0\ , \hspace{.2cm} c_H=\frac{5}{16}\, ,
\end{equation}
while the other coefficients are unaffected. However, the exact $\sigma$-model structure below the scale $m_\rho$ is now lost, since the corrections of order 
$\gsm/g_\rho\sim g_L/g_R$ are important.

\subsection{Little Higgs model with custodial symmetry}
\label{sec:custodialLH}

The littlest Higgs model with custodial symmetry~\cite{Chang:2003zn} is based on the coset $SO(9)/(SO(5)\times SO(4))$ with an $SU(2)_L\times SU(2)_R \times SU(2) \times U(1)$ subgroup gauged ($g_L, g_R, g_2$ and $g_1$ are the respective gauge couplings). This Little Higgs model will be described, below the mass of the new resonances, by a SILH Lagrangian with a $SO(5)/SO(4)$ structure. 

The global symmetry breaking of this Little Higgs model is conveniently parametrized by a symmetric representation of $SO(9)$ taking a vev of the form
\begin{equation}
\langle \Sigma \rangle = 
\left(
\begin{array}{ccc}
& & \mathbf{1}_4\\
& 1\\
\mathbf{1}_4 \\
\end{array}
\right) .
\end{equation}
Among the 20 Goldstone bosons, 6 are eaten in the gauge symmetry breakings $SU(2)_L \times SU(2)\to SU(2)_W$ and $SU(2)_R\times U(1) \to U(1)_Y$ and we are left with a charged doublet $H_{1/2}$, a neutral triplet $\phi_{0}$, a charged triplet  $\phi_{1}$ and a neutral singlet $s_0$ (the subscripts denote the $U(1)_Y$ charges of the fields). A convenient parametrization of these Goldstones is
\begin{equation}
	\label{eq:PGBSO9}
\Sigma= e^{i\Pi/f}\, \langle \Sigma \rangle\,  e^{i\Pi^T /f} = e^{2i\Pi/f} \langle \Sigma \rangle
\hspace{.5cm} \mathrm{with}\hspace{.5cm} 
2i\Pi = 
\left(
\begin{array}{ccc}
0_{4} & H/\sqrt{2} & -\phi/2\\
- H^T/\sqrt{2} & 0 & H^T/\sqrt{2}\\
\phi/2 & -H/\sqrt{2} & 0_{4} \\
\end{array}
\right).
\end{equation}
$H$ is the real 4-component vector corresponding to the Goldstone doublet, while
$\phi$ is a real $4\times 4$ symmetric matrix and it contains the singlet and the triplets.
The kinetic term of the $\Sigma$ field generates the interactions among the Goldstones
\begin{equation}
\mathcal{L} = \frac{f^2}{4} \tr D_\mu \Sigma^\dagger D^\mu \Sigma .
\end{equation}
Under the unbroken $SU(2)_W\times U(1)_Y$, the heavy vector fields transform as a neutral triplet, a neutral singlet and a charged singlet whose masses are
\begin{equation}
m_{W_H}^2=(g_2^2+g_L^2) f^2 ,\ m_{B_H}^2=(g_1^2+g_R^2)f^2, 
\ m_{A_H^\pm}^2= g_R^2 f^2 .
\end{equation}
The gauge couplings of the unbroken gauge symmetries are given by the usual formulae
\begin{equation}
\frac{1}{g^2}=\frac{1}{g_L^2}+\frac{1}{g_2^2}~, ~~~~~~~~ \frac{1}{{g'}^2}=\frac{1}{g_R^2}+\frac{1}{g_1^2}~.
\end{equation}

By the construction of the model, when the $SU(2)$ and $U(1)$ gauge couplings are turned off,  the gauging of $SU(2)_L$ and $SU(2)_R$ respects a $SO(5)$ global symmetry whose breaking to $SO(4)$ leaves the $H$ doublet as exact Goldstone bosons while the triplets and the singlet acquire a mass of order $m_\rho \equiv g_\rho f$ (we have considered $g_L=g_R\equiv g_\rho$ for concreteness). Hence, the $SO(5)/SO(4)$ $\sigma$-model structure below the scale $m_\rho$, obtained after integrating out the $SU(2)_L \times SU(2)_R$ gauge fields that do not couple to fermions and integrating out  as well the heavy triplet and scalar, which amounts to the constraint
\begin{equation}
\phi = \frac{HH^T}{2f}.
\end{equation}
The oblique corrections are found to be
\begin{equation}
{\widehat S} =  \frac{m_W^2}{m_\rho^2},\hspace{.2cm}
{\widehat T} = 0,\hspace{.2cm}
W= \frac{g^2}{g_\rho^2} \frac{m_W^2}{m_\rho^2},\hspace{.2cm}
Y=  \frac{g'^2}{g_\rho^2} \frac{m_W^2}{m_\rho^2},
\end{equation}
which allow us to identify the coefficients of the effective Lagrangian 
\begin{equation}
c_W + c_B =1,\ c_T=0, \ c_{2W}=1 \ \mathrm{and}\ c_{2B}=1.
\end{equation}
The value of $c_H$ can be computed, exactly as before, by looking at the $hWW$ coupling at $p^2=0$.
We obtain
\begin{equation}
c_H= \frac 12 .
\end{equation}
The factor 2 of disagreement with the value, eq.~(\ref{ch}), of $c_H$ computed in the Holographic Higgs model simply fixes the relative normalization of the decay constants in the two models.

\section{Phenomenology of SILH}
\label{secphen}

In this section we analyze the effects of the SILH interactions and study how they can be tested at future colliders.  Let us start by considering the new interaction terms involving the physical Higgs boson. For simplicity, we work in the unitary gauge and write the SILH effective Lagrangian in~\eq{lsilh} only for the real Higgs field $h$ (shifted such that $\vev{h} =0$). We reabsorb the contributions from $c_f$ and $c_6$ to the SM input parameters
(fermion masses $m_f$,
Higgs mass $m_H$, and vacuum expectation value $v=246\GeV$). Similarly, we redefine the gauge fields and the gauge coupling constants  and we make the gauge kinetic terms canonical. In this way, the SILH effective Lagrangian is composed by the usual SM part (${\cal L}_{\rm SM}$), written in terms of the usual SM input parameters (physical masses and gauge couplings),
by new Higgs interactions (${\cal L}_h$), and 
new interactions involving only gauge bosons  (${\cal L}_V$)  which, at leading order, are given by
\bea
&&{\cal L}_h = \xi\, \left\{ \frac{c_H}{2}\left(1+\frac{h}{v}\right)^2 \partial^\mu h\partial_\mu h 
-c_6 \frac{m_H^2}{2v^2} \left( vh^3+\frac{3h^4}{2}+\dots \right) 
+c_y \frac{m_f}{v} {\bar f}f \left( h+\frac{3h^2}{2v}+\dots \right)
\right. \nonumber \\  &&
+ \left( \frac{h}{v}+\frac{h^2}{2v^2}\right) \left[ \frac{g^2}{2g_\rho^2} 
\left( {\hat c}_W W^-_\mu {\cal D}^{\mu \nu} W^+_\nu +{\rm h.c.} \right)
+\frac{g^2}{2g_\rho^2} Z_\mu {\cal D}^{\mu \nu} \left[ {\hat c}_Z Z_\nu +
\left( \frac{2 {\hat c}_W}{\sin 2\theta_W}-\frac{{\hat c}_Z}{\tan\theta_W}\right) A_\nu \right]
\right. \nonumber \\ &&
 -\frac{ g^2}{(4\pi)^2} \left( \frac{c_{HW}}{2}
W^{+\mu \nu} W^-_{\mu \nu} +\frac{c_{HW}+\tan^2\theta_W c_{HB}}{4} Z^{\mu \nu}Z_{\mu \nu} -2 \sin^2\theta_W c_{\gamma Z}F^{\mu \nu}Z_{\mu \nu} \right) + \dots 
\nonumber \\ &&\left. \left. 
+\frac{\alpha g^2 c_\gamma}{4\pi g_\rho^2} F^{\mu \nu}F_{\mu \nu} +\frac{\alpha_s y_t^2 c_g}{4\pi g_\rho^2} G^{a \mu \nu}G^a_{\mu\nu}
\right] \right \}
\label{leff}
\eea
\beq
{\hat c}_W=c_W+ \left( \frac{g_\rho}{4\pi}\right)^2 c_{HW}
\eeq
\beq
{\hat c}_Z={\hat c}_W+\tan^2\theta_W \left[ c_B +\left( \frac{g_\rho}{4\pi}\right)^2  c_{HB}\right]
\eeq
\beq
c_{\gamma Z}=\frac{c_{HB}-c_{HW}}{4\sin 2\theta_W}
\eeq
\bea
{\cal L}_V&=&-\frac{\tan\theta_W}{2} {\widehat S}\, W^{(3)}_{\mu\nu}B^{\mu\nu}-i g\cos\theta_W g_1^Z Z^\mu \left( W^{+\nu} W^-_{\mu\nu} - W^{-\nu} W^+_{\mu\nu} \right) \nonumber \\&&
 -ig \left( \cos\theta_W \kappa_Z Z^{\mu\nu} +\sin\theta_W \kappa_\gamma A^{\mu\nu}\right) W^+_\mu W^-_\nu \label{ellev}
\eea
\beq
{\widehat S}=\frac{m_W^2}{m_\rho^2}\left (c_W+c_B\right) ,~~~~~g_1^Z=\frac{m_Z^2}{m_\rho^2}~{\hat c}_W \eeq
\beq
\kappa_\gamma = \frac{m_W^2}{m_\rho^2}\left( \frac{g_\rho}{4\pi}\right)^2 \left( c_{HW}+ c_{HB}\right)  ,~~~~~ \kappa_Z = g_1^Z
-\tan^2\theta_W \kappa_\gamma.
\eeq
In ${\cal L}_V$ we have included only trilinear terms in gauge bosons and dropped the effects of $O_{2W}$,  $O_{2B}$, $O_{3W}$. In ${\cal L}_h$
we have kept only the first powers in the Higgs field $h$ and the gauge fields. We have defined $W^\pm_{\mu \nu} =\partial_\mu W_\nu^\pm -\partial_\nu W^\pm_\mu$ (and similarly for the $Z_\mu$ and the photon $A_\mu$) and
${\cal D}_{\mu \nu} =\partial_\mu \partial_\nu - \Box g_{\mu \nu}$. Notice that for on-shell
gauge bosons ${\cal D}_{\mu \nu}A^{\mu i}=M_i^2 A_\nu^i$. Therefore ${\hat c}_W$ and ${\hat c}_B$
generate a Higgs coupling to gauge bosons which is proportional to mass, as in the SM, and do not generate any Higgs coupling to photons. Notice also that the corrections to trilinear vector boson vertices satisfy the relation $g_1^Z=k_Z+\tan^2\theta_W k_\gamma$ \cite{Kuroda:1986ci}.

The new interactions in ${\cal L}_h$, see~\eq{leff}, modify the SM predictions for Higgs production and decay.  At quadratic order in $h$, the coefficient $c_H$ generates an extra contribution to the Higgs kinetic term. This can be reabsorbed by redefining the Higgs field according to $h \to h/ \sqrt{1+\xi c_H}$
(see  appendix~B for an alternative redefinition of the Higgs  that  removes  the derivative terms  of  the Higgs -- first term of~\eq{leff}).
The effect of $c_H$ is then to
renormalize by a factor $1-\xi c_H/2$,  with respect to their SM value, the couplings of the canonical field $h$  to all other fields. Notice that the Higgs field redefinition also shifts the value of $m_H$ (but not of $v$ or $m_f$).

We can express the modified Higgs couplings in terms of the decay widths in units of the SM prediction, expressed in terms of physical pole masses (for a review of the Higgs properties in the SM, see~\cite{Djouadi:2005gi},
\beq
\Gamma \left( h\to f\bar f \right)_{\rm SILH} = \Gamma \left( h\to f\bar f \right)_{\rm SM} 
\left[ 1-  \xi \left( 2c_y + c_H \right) \right]
\label{gam1}
\eeq
\beq
\Gamma \left( h\to W^+W^- \right)_{\rm SILH} = \Gamma \left( h\to W^+W^{(*)-} \right)_{\rm SM} 
\left[ 1- \xi \left( c_H -\frac{g^2}{g_\rho^2} {\hat c}_W\right) \right]
\label{gam2}
\eeq
\beq
\Gamma \left( h\to ZZ \right)_{\rm SILH} = \Gamma \left( h\to ZZ^{(*)} \right)_{\rm SM} 
\left[ 1- \xi \left( c_H -\frac{g^2}{g_\rho^2} {\hat c_Z}   \right) \right]
\label{gam3}
\eeq
\beq
\Gamma \left( h\to gg \right)_{\rm SILH} = \Gamma \left( h\to gg \right)_{\rm SM} 
\left[ 1- \xi ~{\rm Re}\left( 2c_y + c_H +\frac{4y_t^2c_g}{g_\rho^2I_g}\right) \right]
\label{gam4}
\eeq
\beq
\Gamma \left( h\to \gamma \gamma \right)_{\rm SILH} = \Gamma \left( h\to \gamma \gamma \right)_{\rm SM} 
\left[ 1- \xi ~{\rm Re}\left( \frac{2c_y+c_H}{1+J_{\gamma}/I_{\gamma}} + \frac{c_H -\frac{g^2}{g_\rho^2} {\hat c}_W}{1+I_{\gamma}/J_{\gamma}}+\frac{\frac{4g^2}{g_\rho^2}c_\gamma}{I_{\gamma}+J_{\gamma}}\right) \right]
\label{gam5}
\eeq
\beq
\Gamma \left( h\to \gamma Z \right)_{\rm SILH} = \Gamma \left( h\to \gamma Z \right)_{\rm SM} 
\left[ 1- \xi ~{\rm Re} \left( \frac{2c_y+c_H}{1+J_{Z}/I_{Z}} + \frac{c_H -\frac{g^2}{g_\rho^2} {\hat c}_W}{1+I_{Z}/J_{Z}}+\frac{4c_{\gamma Z}}{I_{Z}+J_{Z}}\right) \right] .
\label{gam6}
\eeq
Here we have neglected in $\Gamma ( h\to W^+W^- ,ZZ)_{\rm SILH}$ the subleading effects from $c_{HW}$ and $c_{HB}$, which are parametrically smaller than a SM one-loop contribution. The loop functions $I$ and $J$ are given in appendix~C. 

The leading effects on Higgs physics, relative to the SM, come from the three coefficients $c_H$, $c_y$, $c_{\gamma Z}$, although $c_{\gamma Z}$ has less phenomenological relevance since it affects only the decay $h\to \gamma Z$. The rules of SILH select the operators proportional to $c_H$ and $c_y$ as the most important ones for LHC studies, as opposed to totally model-independent operator analyses~\cite{Manohar:2006gz,thaler,Hankele:2006ma} which often lead to the conclusion that the dominant effects should appear in the vertices $h\gamma \gamma$ and $hgg$, since their SM contribution occurs only at loop level. Therefore, we believe that an important experimental task to understand the nature of the Higgs boson will be the extraction of $c_H$ and $c_y$ from precise measurements of the Higgs production rate ($\sigma_h$) and branching ratios ($BR_h$).
The contribution from $c_H$ is universal for all Higgs couplings and therefore it does not affect the Higgs branching ratios, but only the total decay width and the production cross section. The measure of the Higgs decay width at the LHC is very difficult and it can be reasonably done only for rather heavy Higgs bosons, well above the two gauge boson threshold, while the spirit of our analysis is to consider the Higgs as a pseudo-Goldstone boson, and therefore relatively light. However, for a light Higgs, LHC experiments can measure the product $\sigma_h \times BR_h$ in many different channels: production through gluon, gauge-boson fusion, and top-strahlung; decay into $b$, $\tau$, $\gamma$ and (virtual) weak gauge bosons.
At the LHC with about 300~fb$^{-1}$, it is possible to measure Higgs production rate times branching ratio in the various channels
 with 20--40~\% precision~\cite{coup}, although a determination of the $b$ coupling is quite challenging~\cite{incand}.
 This will translate into a sensitivity on $|c_H \xi |$ and $|c_y \xi |$ up to 0.2--0.4.

\begin{figure}[t!]
\begin{center}
\includegraphics[width=12cm]{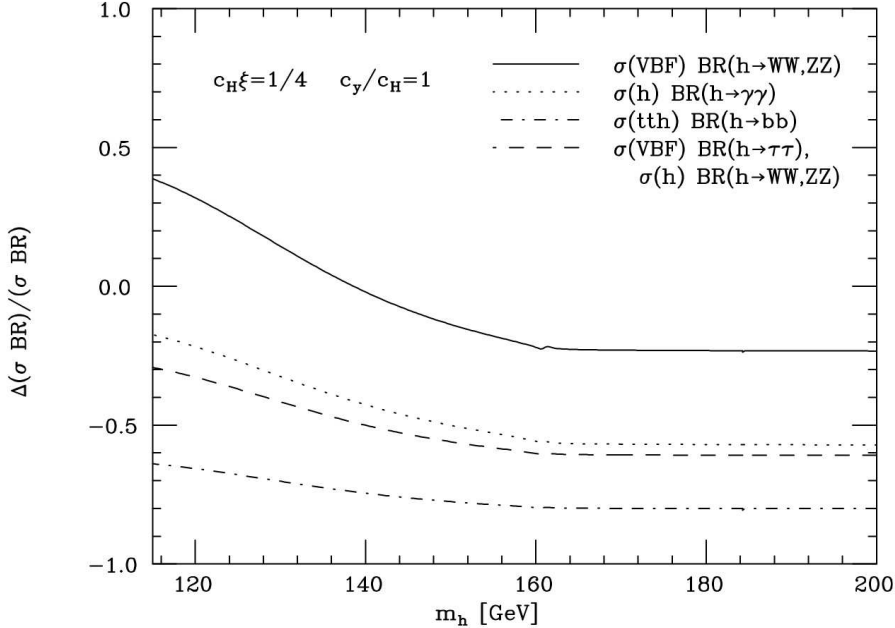}
\end{center}
\caption{The deviations from the SM predictions of Higgs production cross sections ($\sigma$) and decay branching ratios ($BR$) defined as $\Delta (\sigma ~BR)/(\sigma ~BR)=(\sigma ~BR)_{\rm SILH}/(\sigma ~BR)_{\rm SM} -1$. The predictions are shown for some of the main Higgs discovery channels at the LHC with production via vector-boson fusion (VBF), gluon fusion ($h$), and topstrahlung ($tth$). The SILH Lagrangian parameters are set by $c_H\xi =1/4$, $c_y/c_H=1$ and we have included also the terms quadratic in $\xi$, not explicitly shown in eqs.~(\ref{gam1})--(\ref{gam6}).
 }
\label{fig1}
\end{figure}

In fig.~\ref{fig1}, we show our prediction for the relative deviation from the SM expectation in the main channels for Higgs discovery at the LHC, in the case $c_H\xi =1/4$ and $c_y/c_H=1$ (as in the Holographic Higgs). For $c_y/c_H=0$, the deviation is universal in every production channel and is given by $\Delta (\sigma ~BR)/(\sigma ~BR)=-c_H\xi$.
 
Cleaner experimental information can be extracted from ratios between the rates of processes with the same Higgs production mechanism, but different decay modes. In measurements of these ratios of decay rates, many systematic uncertainties drop out. Our leading-order ($g_\rho \gg \gsm$) prediction is that $\Delta [\Gamma (h\to ZZ)/\Gamma ( h\to W^+W^-)] =0$, $\Delta [\Gamma (h\to f\bar f)/\Gamma ( h\to W^+W^-)] =-2\xi c_y$, $\Delta [\Gamma (h\to \gamma \gamma)/\Gamma ( h\to W^+W^-)] =-2\xi c_y(1+J_\gamma /I_\gamma )^{-1}$.
However, the Higgs coupling determinations at the LHC will still be limited by statistics, and therefore they can benefit from a luminosity upgrading, like the SLHC. At a linear collider, like the ILC, precisions on $\sigma_h \times BR_h$ can reach the percent level~\cite{ilc}, providing a very sensitive probe on the new-physics scale. Moreover, a linear collider can test the existence of $c_6$, since the triple Higgs coupling can be measured with an accuracy of about 10\% for $\sqrt{s} =500$~GeV and an integrated luminosity of 
1000 fb$^{-1}$~\cite{Barger:2003rs}.

Deviations from the SM predictions of Higgs production and decay rates, could be a hint towards models with strong dynamics, especially if no new light particles are discovered at the LHC. However, they do not unambiguously  imply the existence of a new strong interaction. The most characteristic signals of a SILH have to be found in the very high-energy regime. Indeed, a peculiarity of SILH is that, in spite of the light Higgs, longitudinal gauge-boson scattering amplitudes grow with energy and the corresponding interaction becomes strong, eventually violating tree-level unitarity at the cutoff scale. Indeed, the extra Higgs kinetic term proportional to $c_H \xi$ in~\eq{leff} prevents Higgs exchange diagrams  from accomplishing the exact cancellation, present in the SM, of the terms growing with energy in the amplitudes. Therefore, although the Higgs is light, we obtain strong $WW$ scattering at high energies.

From the operator $O_H\equiv  \partial^\mu ( H^\dagger H ) \partial_\mu ( H^\dagger H ) 
$ in~\eq{lsilh}, using the equivalence theorem~\cite{chan}, it is easy to derive the following high-energy limit of the scattering amplitudes for longitudinal gauge bosons
\beq
{\cal A}\left( Z^0_L Z^0_L \to W^+_L W^-_L \right) =
{\cal A}\left( W^+_L W^-_L \to Z^0_L Z^0_L \right)=
-{\cal A}\left( W_L^\pm W_L^\pm \to W_L^\pm W_L^\pm \right)=\frac{c_H s}{f^2},
\label{strsc}
\eeq
\beq
{\cal A}\left( W^\pm Z^0_L \to W^\pm Z^0_L\right)=\frac{c_H t}{f^2},~~~~
{\cal A}\left( W^+_L W^-_L \to W^+_L W^-_L \right) =\frac{c_H (s+t)}{f^2},
\label{strsc2}
\eeq
\beq
{\cal A}\left( Z^0_L Z^0_L \to Z^0_L Z^0_L \right) =0.
\label{strsc3}
\eeq
This result is correct to leading order in $s/f^2$, and to all orders in $\xi$ in the limit $\gsm=0$, when the $\sigma$-model is exact. The absence of corrections in $\xi$ follows from the non-linear symmetry of the $\sigma$-model, corresponding to the action of the generator $T_h$, associated with the neutral Higgs, under which $v$ shifts.
Therefore we expect that corrections can arise only at $\mathcal{O} (s/m_\rho^2)$. The growth with energy of the amplitudes in eqs.~(\ref{strsc})--(\ref{strsc3}) is strictly valid only up to the maximum energy of our effective theory, namely $m_\rho$. The behaviour above $m_\rho$ depends on the specific model realization. In the case of the Little Higgs, we expect that the amplitudes continue to grow with $s$ up to the cut-off scale $\Lambda$.   In 5D models, like the Holographic Goldstone, the growth of the elastic amplitude is softened by KK exchange, but the inelastic channel dominate and strong coupling is reached at a scale $\sim 4\pi m_\rho/g_\rho$.
Notice that the result in eqs.~(\ref{strsc})--(\ref{strsc3}) is exactly proportional to the scattering amplitudes obtained in a Higgsless SM~\cite{chan}. Therefore, in theories with a SILH, the cross section at the LHC for producing longitudinal gauge bosons with large invariant masses can be written as
\beq
\sigma \left( pp \to V_L V^\prime_L X\right)_{c_H} =
\left( c_H \xi \right)^2 \sigma \left( pp \to V_L V^\prime_L X\right)_{\not \! H},
\eeq
where $\sigma ( pp \to V_L V^\prime_L  X)_{\not \! H}$ is the cross section in the SM without Higgs, at the leading order in $s/(4\pi v)^2$.
With about 200~fb$^{-1}$ of integrated luminosity, it should be possible to identify the signal of a Higgsless SM with about 30--50\% accuracy~\cite{han,Butterworth:2002tt,Accomando:2006vj}. This corresponds to a sensitivity up to
$c_H \xi \simeq 0.5$--0.7.

In the SILH framework, the Higgs is viewed as a pseudo-Goldstone boson and therefore its properties are directly related to those of the exact (eaten) Goldstones, corresponding to the longitudinal gauge bosons. Thus, a generic prediction of SILH is that the strong gauge boson scattering is accompanied by strong production of Higgs pairs. Indeed we find that, as a consequence of the $O(4)$ symmetry of the $H$ multiplet, the amplitudes for Higgs pair-production grow with the center-of-mass energy as~\eq{strsc},
\beq
{\cal A}\left( Z^0_L Z^0_L \to hh \right) =
{\cal A}\left( W^+_L W^-_L \to hh \right)=\frac{c_H s}{f^2}.
\label{strsc4}
\eeq
Notice that scattering amplitudes involving longitudinal gauge bosons and a single Higgs vanish. This is a consequence of the $Z_2^4$ parity embedded in the $O(4)$ symmetry of the operator $O_H$, under which each Goldstone change sign. Non-vanishing amplitudes necessarily involve an even number of each species of Goldstones.

Using eqs.~(\ref{strsc}), (\ref{strsc2}) and (\ref{strsc4}), we can relate the Higgs pair production rate at the LHC to the longitudinal gauge boson cross sections
\beq
2\sigma_{\delta ,M} \left( pp \to hh X\right)_{c_H} = 
\sigma_{\delta ,M} \left( pp \to W^+_L W^-_L X\right)_{c_H}
+\frac 16 \left( 9- \tanh^2 \frac \delta 2 \right)
\sigma_{\delta ,M} \left( pp \to Z^0_L Z^0_L X\right)_{c_H}.
\label{sumr}
\eeq
Here  all cross sections $\sigma_{\delta ,M} $ are computed with a cut on the pseudorapidity separation between the two final-state particles (a boost-invariant quantity) of $|\Delta \eta |<\delta$,
and with a cut on the two-particle invariant mass $\hat s >M^2$. The sum rule in~\eq{sumr} is a characteristic of SILH. However, the signal from Higgs-pair production at the LHC is not so prominent. It was suggested that, for a light Higgs, this process is best studied in the channel $b\bar b \gamma \gamma$~\cite{ple}, but the small branching ratio of $h\to \gamma \gamma$ makes the SILH rate unobservable. However, in SILH, one can take advantage of the growth of the cross section with energy. Although we do not perform here a detailed study, it may be possible that, with sufficient luminosity, the signal of $b\bar b b \bar b$ with high invariant masses could be distinguished from the SM background. Notice however that, because of the high boost of the Higgs boson, the $b$ jets are often not well separated. The case in which the Higgs decays to two real $W$'s appears more promising for detection.
The cleanest channel is the one with two like-sign leptons, where
$hh\to \ell^\pm\ell^\pm\nu\nu \,{\rm jets}$, studied in refs. \cite{ple,thaler}.

The operator $O_H$ is purely generated by the strongly-interacting sector, as indicated by its  $\sim  g_\rho^2/m_\rho^2$ coefficient.  Also $c_y$ and $c_6$, even though they arise from the interplay between weak and strong couplings, are sensitive to $g_\rho^2/m_\rho^2$  and thus indirectly test the  non-linearity of the Higgs sector.
 Therefore probing the  effects of these couplings is crucial for testing SILH. 
The operators  $O_{W,B}$, $O_{BB}$ and $O_g$, on the other hand,
only depend on the scale of new physics, not on its strength, as indicated by their $\propto 1/m_\rho^2$  coefficient. Because of this fact,  in the strong coupling limit $g_\rho\gg \gsm$,
their effects in Higgs decay rates are subleading with respect to those induced by $c_H$ and $c_y$. Indeed the contribution of $O_{W,B}$, $O_{BB}$ and $O_g$ to amplitudes scales like $s/m_\rho^2$ times the SM contribution. While at electroweak energies this effect is very small, it can become a sizeable at higher energies.

As an example, the operator $O_{g}$ contributes to processes like $gg \to hh,Z^0_L Z^0_L, W^+_L W^-_L$ with
scattering amplitudes ${\cal A}(gg\to hh )_{\rm SILH} \simeq {\cal A}(gg\to hh )_{\rm SM} c_g s/m_\rho^2$ which are, at most, of the same order of magnitude of the SM one, for the maximal energy $s\simeq m_\rho^2$, where the corresponding resonances can be directly produced. Double-Higgs production at the LHC from $O_{g}$ was recently studied in ref.~\cite{thaler}.
Also the operators $O_{W,B}$ can contribute to high-energy production of Higgs and longitudinal gauge bosons. Indeed, by using the equations of motion for the gauge bosons, these operators can be rewritten as the product of a fermionic current times a bilinear in $H$ and can give new-physics effects in 
$\bar q q^\prime \to h Z^0_L,h W_L^\pm,Z^0_L W_L^\pm , W_L^\pm W_L^\mp$. 

The operators $O_{HW,HB}$, in spite of their overall 1-loop suppression, are sensitive to $g_\rho^2/m_\rho^2$, indicating that in principle they probe the strong dynamics. Indeed,  they induce
corrections to the process $h\to Z\gamma$ and to the magnetic moment anomaly of the $W$ that are  $O(v^2/f^2)$ relative to the SM
contribution. In practice, however, these quantities are experimentally not  well accessible. 
 Therefore $c_{HW}$ and $c_{HB}$ should be tested in vector boson production, where their contribution relative to the SM
is quantitatively similar (indeed even $g_\rho^2/16\pi^2$ smaller) to that of $c_{W,B}$.

The effect of the operators $O_{W,B}$, $O_{HW,HB}$ can be tested through precise measurements of triple gauge vertices. At the LHC, the anomalous couplings in~\eq{ellev} can be probed at the level of $10^{-1}$--$10^{-2}$~\cite{Dobbs:2005ev}, but at a sub-TeV linear collider the precision can be improved up to  $10^{-3}$--$10^{-4}$~\cite{Abe:2001nq,Beyer:2006hx}. This is highly competitive with the $\widehat S$ determination at LEP and can probe values of $m_\rho$ up to 6--8~TeV.

As discussed in sect.~\ref{intro}, the signals studied in this paper are important because they are model-independent tests of a strongly-interacting electroweak-breaking sector in presence of a light Higgs. However, the first evidence for this kind of new physics at the LHC may come from production of the resonances at the mass scale $m_\rho$. Therefore, it is useful to compare the indirect effects studied here with the direct resonance production. As an illustrative example, we consider the case of new spin-one charged resonances $\rho_H^\pm$. These particles can be interpreted as bound states analogous to the technirho, in composite models, or as the heavy gauge bosons $W_H^\pm$ in Little-Higgs models. They have mass $m_\rho$, coupling  to the strong sector (Higgs and longitudinal gauge bosons) equal to $g_\rho$ and coupling to the weak sector (quarks, leptons and transverse gauge bosons) equal to $g^2/g_\rho$. Indeed, in the effective theory below $m_\rho$, they give rise to the operators $O_{W,B}$ with coefficients of order $g^2/m_\rho^2$. The
cross section for the resonant production of $\rho_H^+$ is
\beq
\sigma \left( pp\to \rho_H^++X\right) = \frac{\pi g^4}{12 g_\rho^2} \left. \frac{\tau d{\cal L}}{{\hat s} d \tau}\right|_{\hat s =m_\rho^2},
\eeq
where $\tau /\hat s d{\cal L}/d\tau$ is the parton luminosity at an energy equal to the resonance mass. For $2~{\rm TeV}\lsim m_\rho \lsim 4~{\rm TeV}$, we find
\beq
\sigma \left( pp\to \rho^\pm_H+X\right) = \left( \frac{4\pi}{g_\rho}\right)^2 \left( \frac{3~{\rm TeV}}{m_\rho}\right)^{6}~0.5~{\rm fb}.
\label{respit}
\eeq

The $\rho_H^\pm$ branching ratios are
\beq
BR\left( \rho_H^- \to \mu \bar \nu \right) =\frac{1}{3} BR\left( \rho_H^- \to b \bar t \right) =\frac{2g^4}{g_\rho^4} \left( 1+ \frac{24g^4}{g_\rho^4}\right)^{-1} ,
\label{bruno}
\eeq
\beq
BR\left( \rho_H^- \to hW^-\right) = BR\left( \rho_H^- \to Z^0W^-\right)=\frac 12\left( 1+ \frac{24g^4}{g_\rho^4}\right)^{-1}.
\label{brdue}
\eeq
The resonances are most easily detected when they decay directly into leptons or top quarks. However, as shown in eqs.~(\ref{bruno})--(\ref{brdue}),
for large $g_\rho$, these decay modes are suppressed and  gauge and Higgs bosons then provide the dominant decay channels (in some specific models the coupling 
of $\rho_H$ to the top  can be larger than $g^2/g_\rho$  -- see section~\ref{secmod}). 
Notice that, as $g_\rho$ grows, the experimental identification of the resonance becomes increasingly hard, not only because the leptonic signal is suppressed, but also because the decay width becomes large. Detection of a broad resonance decaying into gauge and Higgs bosons is experimentally challenging and the study of indirect signals becomes more important in the region of large $g_\rho$.

 For order-unity coefficients $c_i$, we have described the  SILH 
 in terms of the two parameters $m_\rho$ and $g_\rho$. An alternative description cane be done in terms of two mass scales. They  can be chosen as $4\pi f$, the scale at which the $\sigma$-model would become fully strongly-interacting in the absence of new resonances, and $m_\rho$, the scale at which new states appear. An upper bound on $m_\rho$ is obtained from the theoretical NDA requirement $m_\rho <4\pi f$, while a lower bound on $m_\rho$ comes from the experimental constraint on the $\widehat S$ parameter, see~\eq{eq:hatS}. 

Searches at the LHC, and possibly at the ILC, will probe unexplored regions of the $4\pi f$--$m_\rho$ space. Precise measurements of Higgs production and decay rates at the LHC will be able to explore values of $4\pi f$ up to 5--7~TeV, mostly testing the existence of $c_H$ and $c_y$. These measurements can be improved with a luminosity upgrading of the LHC. Higgs-physics studies at a linear collider could reach a sensitivity on $4\pi f$ up to about 30~TeV. Analyses of strong gauge-boson scattering and double-Higgs production at the LHC can be sensitive to values of $4\pi f$ up to about 4~TeV. These studies are complementary to Higgs precision measurements, as they test only the coefficient $c_H$ and probe processes highly characteristic of a strong electroweak-breaking sector with a light Higgs boson.

On the other side, the parameter $m_\rho$ can be probed at colliders by studying pair-production of longitudinal gauge bosons and Higgs, by testing triple gauge vertices or, more directly, by producing the new resonances. For fixed $m_\rho$, resonance production at the LHC will overwhelm the indirect signal of longitudinal gauge boson and Higgs production, at large $4\pi f$ (small $g_\rho$). However, at low $4\pi f$ (large $g_\rho$) resonance searches become less effective in constraining the parameter $m_\rho$ and the indirect signal gains importance. 
While the search for new resonances is most favorable at the LHC, precise measurements of triple gauge vertices at the ILC can test $m_\rho$ up to 6-8~TeV. With complementary information from collider data,  we will explore a large portion of the interesting region of the $4\pi f$--$m_\rho$ plane, testing the composite nature of the Higgs.

\section{Strongly-interacting  top quark}
\label{sectop}

In sect.~\ref{secmod} we have seen that, in some explicit realizations of  the SILH, 
 the top  quark is required to be strongly coupled to the 
resonances of the electroweak-breaking sector.
Here we want to     study,  in a model-independent way,  the  phenomenological implications of this  strongly-coupled top quark, much in the same spirit of
sect.~\ref{secstruc} for   the  case of the Higgs boson.

Let us  first consider the case in which, in addition to   the Higgs, 
     the right-handed top  also belongs to the strongly-coupled  sector.
The low-energy effective Lagrangian can be written by generalizing the rules
1, 2 and 3 of sect.~\ref{secrules}, noticing that each $t_R$ leg added to leading interactions carries an extra factor $1/(fm_\rho^{1/2})$. We find three 
 dimension-6 operators suppressed by   $1/f^2$  and involving $t_R$: 
\begin{equation} 
\frac{c_t y_t}{f^2} H^\dagger H \bar q_L \tilde Ht_R+h.c.
+\frac{ic_R}{f^2}  H^\dagger D_\mu H \bar t_R\gamma^\mu t_R+
\frac{c_{4t}}{f^2}  (\bar t_R\gamma^\mu t_R)(\bar t_R\gamma_\mu t_R)
\, .
\label{tr}
\end{equation}
We are not considering  dimension-6 operators suppressed by $1/m_\rho^2$
since their effects are smaller than those in eq.~(\ref{tr}) for large $g_\rho$.
The first term of eq.~(\ref{tr}) 
was already included in eq.~(\ref{lsilh}). Nevertheless,   here
it is only present for the top quark and therefore it violates   the universality 
of  $c_y$. The difference $c_t-c_y$ can be viewed as originating from an insertion of $H^\dagger H /f^2$ on the $t_R$ line.
The second term of eq.~(\ref{tr})    violates the custodial symmetry,
and therefore  it  generates a contribution to $\widehat T$ at  the one-loop level
\begin{equation} 
\widehat T\sim  \frac{N_cc_R^2v^2\Lambda^2}{16\pi^2 f^4}=
0.02\, c_R^2\left(\frac{\Lambda}{f}\right)^2\xi\, ,
\label{estimate}
\end{equation}
where  $\Lambda$ is the scale
that cuts off the one-loop momentum divergence. 
In models in which  $\Lambda\sim m_\rho$ the 95\% CL bound 
$\widehat T\lsim 0.002$ translates, via eq.~(\ref{estimate}),
 into a   severe  upper bound on $c_R^2\xi$. 
 This bound on $c_R$ can be easily  satisfied in models
  in which  the strong sector preserves a custodial symmetry under which 
  $t_R$  transforms as a  singlet.
This guarantees  
  $c_R=0$ at tree-level.
Another possibility to  evade  the bound on $c_R$ is to reduce the scale
 $\Lambda$ in \eq{estimate}.  
This can be achieved  in  models   in which $t_R$ transforms non-trivially
under the custodial group
as discussed in sect.~\ref{secmod}.
In this case $\Lambda\sim m_{cust}$ 
where $m_{cust}$ is the mass of   the  custodial partners of the $t_R$.
Assuming
 $m_{cust}\ll m_\rho$  we can satisfy the bound from $\widehat T$ 
even if $c_R\sim 1$.

Similarly, we can   consider the case in which $t_L$ and $H$  are strongly coupled.
We have now the following    $1/f^2$ dimension-6 operators 
in the low-energy Lagrangian involving $q_L=(t_L,b_L)$:
\begin{eqnarray} 
&&\frac{c_q y_b}{f^2} H^\dagger H \bar q_L H b_R+
\frac{c_q y_t}{f^2} H^\dagger H\bar q_L \tilde H t_R+h.c. 
+\frac{ic^{(1)}_L}{f^2}  H^\dagger D_\mu H \bar q_L\gamma^\mu q_L
\nonumber\\
&&
+\frac{ic^{(3)}_L}{f^2}  H^\dagger \sigma^i D_\mu H \bar q_L\gamma^\mu \sigma^i q_L
+\frac{c_{4q}}{f^2}  (\bar q_L\gamma^\mu q_L)(\bar q_L \gamma_\mu q_L)
\, .
\label{opetL}
\end{eqnarray}
The possibility of having a strongly-coupled $q_L$ has, however, severe constraints
from flavor physics due to  $b_L$.
For example, the operator proportional to $c_{4q}$ in eq.~(\ref{opetL}) 
contributes to $\Delta m_B$, the mass difference of neutral $B$ mesons
\begin{equation}
\Delta m_B =\frac 23 \xi c_{4q} m_B \frac{f_B^2}{v^2} \theta_{bd}^2\, ,
\end{equation}
where  the angle $\theta_{bd}$ 
parametrizes   the projection of $b_L$   into  the $d$ mass eigenstate.
From the requirement that the new contribution to $\Delta m_B$ does not exceed 20\% of the experimental value, we obtain
\begin{equation}
\xi c_{4q}\left( \frac{\theta_{bd}}{V_{ub}}\right)^2 < 2\times10^{-3}\, .
\label{bcq}
\end{equation}
Therefore, unless there 
is a flavor-symmetry reason for having an alignment of $b_L$ with a mass eigenstate more accurate than the corresponding CKM angle $V_{ub}$, 
the  bound in eq.~(\ref{bcq}) disfavors a strongly-coupled $q_L$.  
Thus, we will not further consider this possibility.  

Flavor constraints have been studied in detail, {\it e.g.} see~\cite{Agashe:2004cp}, within the framework of warped extra dimensions and they apply to holographic Higgs models whose low energy effective description is described by our Lagrangian~(\ref{lsilh})

\subsection{Phenomenology of a  strongly-interacting  $t_R$}

The presence of the  operators (\ref{tr})
gives non-universal modifications to   the couplings of the top  to the Higgs  and 
gauge bosons.
In particular, we find that 
the couplings  $h\bar tt$  and $Z\bar t_Rt_R$ are given by
\begin{eqnarray} 
g_{htt}&=&\frac{gm_t}{2m_W}\left[1-\xi\left(c_t+c_y+\frac{c_H}{2}\right)\right]\, ,\nonumber\\
g_{Zt_Rt_R}&=&-\frac{2g\sin^2\theta_W}{3\cos\theta_W}
\left(1-\frac{3}{8\sin^2\theta_W}c_R\xi\right)\, .
\end{eqnarray}
At the LHC the coupling $g_{htt}$
can be measured in the process $gg\rightarrow \bar t th,~h\to \gamma \gamma$. An accuracy on $g_{htt}$ up to    $5$\% can be reached
at  a linear  collider with 
$\sqrt{s}=800$~GeV and $L=1000$~fb$^{-1}$~\cite{ilc}.   
The coupling $Zt_Rt_R$  can only be measured with accuracy at future
$e^+e^-$  colliders. For 
$\sqrt{s}=500$~GeV and $L=300$~fb$^{-1}$,  one can reach a 
sensitivity up to  $\xi c_R\sim 0.04$~\cite{ilc}.
Deviations on the SM vertex $Zt_Rt_R$  can also be tested in 
 flavor-violating processes. 
For example,  one-loop penguin diagrams, mediated  by the $Z$,
generates the  effective operator
\beq
{\tilde c}_{ij} \bar d^i_L \gamma_\mu   d^j_L ~\bar f \gamma^\mu(Q_f \sin^2\theta_W -T_f P_L) f\, ,
\label{fla}
\eeq
where $Q_f$ and $T_f$ are the electric charge and the third isospin component of the generic fermion $f$,  $P_L$ is the left chiral projector, and 
\beq
{\tilde c}_{ij}=\frac{ c_R\xi \alpha G_F V^*_{ti}V_{tj} m_t^2}{4\sqrt{2}\pi\sin^2\theta_W m_W^2}\ln \frac{m_\rho}{m_t} \, .
\eeq
This operator  contributes to many rare $\Delta F=1$ processes like $B\to X_s \ell^+ \ell^-$, $B\to X_s \bar \nu \nu $, $B_s \to \ell^+ \ell^-$, $K^+\to \pi^+ \bar \nu \nu $, $\epsilon^\prime /\epsilon$, etc. The typical experimental sensitivities or theoretical uncertainties of these processes is no better than 10--20{\%}. To estimate the new-physics contribution, it is useful to express ${\tilde c}_{ij}$ in units of the SM contribution 
\beq
\frac{ {\tilde c}_{ij}}{{\tilde c}^{\rm SM}_{ij}}=c_R\xi f(x_{Wt} ) \ln \frac{m_\rho}{m_t}\, , 
\eeq
where  $x_{Wt}=m_W^2/m_t^2$ and
\beq
f(x_{Wt})=\frac{(1-x_{Wt})^2}{2[1-7x_{Wt}+6x_{Wt}^2-x_{Wt}(3+2x_{Wt})\ln x_{Wt}]}
\simeq 0.34\, .
\eeq
This shows that present limits from flavour physics give only a mild constraint on $c_R\xi$ and, with improved experimental accuracy,  the coefficient $c_R$ 
 can potentially lead to observable signals.
It is also interesting to consider effects of $c_R$  in rare top decays. 
By defining $\theta_{ti}$ as the mixing angle between the current $t_R$ state and the mass eigenstates, we find
\beq
\Gamma (t\to c Z)=\frac{m_t}{2\pi} \left( \frac{\xi c_R g \theta_{tc}}{16 \cos\theta_W}\right)^2 \frac{(1+2x_{Zt})(1-x_{Zt})^2}{x_{Zt}}\, ,
\eeq
where $x_{Zt}=m_Z^2/m_t^2$. This gives a branching ratio
\beq
BR(t\to cZ)=2\times 10^{-4} \left( \xi c_R \right)^2 \left( \frac{\theta_{tc}}{V_{cb}}\right)^2\, .
\eeq
Since the LHC is expected to reach a sensitivity on $BR(t\to cZ)$ of $2\times 10^{-4}$ with 100~fb$^{-1}$, a signal from $c_R$ is possible, but requires a
 mixing angle  $\theta_{tc}$ larger than the corresponding CKM element $V_{cb}$.

Let us finally comment on possible implications of  
the operator proportional to $c_{4t}$.
Analogously to $c_H$ for $WW$ scattering, this operator
induces a  $t\bar t$ scattering  that  grows with energy. 
At the LHC this will give an enhancement of the 
cross-section $pp\rightarrow t\bar t t\bar t$ where a  $t\bar t$ pair 
is produced by the new 4-top interaction.
The coefficient $c_{4t}$ 
 gives also   contributions to flavor processes.
 For example it contributes to $\Delta m_D$, the mass difference of neutral $D$ mesons: 
 \beq
\Delta m_D =\frac 23 \xi c_{4t} m_D \frac{f_D^2}{v^2} \theta_{tc}^2 \theta_{tu}^2 =\xi c_{4t} 
\left( \frac{\theta_{tc}}{V_{cb}}\right)^2 \left( \frac{\theta_{tu}}{V_{ub}}\right)^2 ~2\times 10^{-11}~{\rm MeV}.
\eeq
For mixing angles of the order of the corresponding CKM elements, this prediction is not far from the present experimental bound, which is $\Delta m_D < 4\times 10^{-11}$~MeV.

\section{Conclusions}
\label{secconc}

If the weak scale originates from dimensional transmutation in some new strong sector then the physics of the
Higgs will manifest important deviations with respect to Standard Model expectations.
Technicolor represents the simplest and perhaps most dramatic such possibility: no narrow state can be identified as the Higgs boson. Simple technicolor is however at odds with electroweak precision tests and largely because
of the absence of a light Higgs resonance. Models where, in addition to the three eaten Goldstone bosons, a light
pseudo-Goldstone Higgs appears in the low-energy theory can fare better in electroweak data for two reasons.
On one hand, a light Higgs screens the infrared contribution to $\widehat S$. On the other hand, the vacuum dynamics of the pseudo-Goldstone is determined by extra parameters (SM couplings among them) and therefore one can imagine obtaining $v^2/f^2$ a little bit below 1,  which is enough to suppress the UV contribution $\widehat S\sim (N_{TC}g^2/16\pi^2)(v^2/f^2)$ below the experimental bound. While work has been done in the past on the effective  low-energy description of Higgsless theories like technicolor, less effort has been devoted to the construction of an effective theory for the
pseudo-Goldstone Higgs. This is of course partly justified by the actual existence of specific models which allow for a quantitative description of the resonance sector. Still, we believe that the construction of such an effective theory is an important task, and this has been the primary goal of this paper. 

After its proposal~\cite{georgikaplan} and some work in the eighties~\cite{othercompositeHiggs}, the idea of a pseudo-Goldstone Higgs was recently revived by its realizations in warped compactifications (Holographic Goldstones) and in Little Higgs models.  As we emphasized in this paper, these new models represent weakly-coupled variants
of the original QCD-like proposals. For what pertains low-energy phenomenology, we found it  useful to characterize these
theories in terms of a mass scale $m_\rho$ and a coupling $g_\rho$. In 5D models, these are respectively the Kaluza--Klein mass and coupling. In Little-Higgs models these are the masses and self-couplings that regulate the quadratic divergence
of the Higgs mass. In the limit $g_\rho\to 4 \pi$, these models  coincide with a generic
 strongly-coupled theory~\cite{georgikaplan}. A crucial test of these theories will undoubtedly proceed through the search for new resonances at the LHC.  This is certainly more true at small enough $g_\rho$, where the resonances are narrow. 
However,  these theories also predict important deviations from the SM in the physics of the Higgs. These deviations are  associated to non-renormalizable operators in the low-energy description.
The study 
of these indirect effects should  nicely complement the direct searches, especially  
for  large $g_\rho$ where the  resonances become heavier and broader. Indeed, the leading
dimension-6 operators
have a coefficient  $\sim (g_\rho/m_\rho)^2 $, indicating the relevance of these effects even when the resonances are heavy provided $g_\rho$ is large.
 
Using our simplified description in terms of $(g_\rho, m_\rho)$, we have derived the form of the leading dimension-6 effective Lagrangian. Our description thus encompasses all models with Goldstone Higgs, although
the effective-Lagrangian approach is best motivated in the regime where $g_\rho$ is large. Models
based on 5D are favored to be in this regime, while in Little Higgs models there is more freedom (some couplings are favored to be large, others can be weak). Our effective Lagrangian  is shown in~\eq{lsilh}.
One can distinguish two classes of effects,  the  ``new couplings", which are genuinely sensitive to the new strong force, and the ``form factors" which are  basically sensitive to the spectrum. 
The ``new couplings'', described by $c_H$, $c_T$, $c_6$ and $c_y$, are determined by an expansion in the Higgs field, and test its strong self interaction, characterized by  $g_\rho /m_\rho=1/f$. The rest of the terms in~\eq{lsilh},
can be basically viewed as higher-derivative dressings of the quadratic (free) Higgs action: as such they do not test equally well the strongly-coupled nature of the Higgs. More precisely, the operators proportional to $c_W$, $c_B$, $c_{\gamma}$ and $c_{g}$ represent genuine ``form factor'' effects since they  have a $1/m_\rho^2$ coefficient. Remarkably,  the coefficients $c_\gamma$ and $c_{g}$ are associated to a suppression factor $\sim (\gsm^2/16\pi^2)(1/m_\rho^2)$ that can be seen as arising from the product of a strong loop $1/(16 \pi^2 f^2)$ factor times $\gsm^2/g_\rho^2$.
This second  factor, which determines the dependence on just $m_\rho = g_\rho f$, and suppresses these effects at large $g_\rho$, is dictated by the Goldstone symmetry and by its preservation by both the gluon and photon fields.
The operators $O_{HW}$ and $O_{HB}$   are special in that they have a structure similar to the form factors but a coefficient $g_\rho^2/(16 \pi^2 m_\rho^2)$ that depends on $g_\rho$, like for the ``new couplings''. In practice, however, they are experimentally less relevant than the ``new couplings"  at measuring $g_\rho/m_\rho$.   Therefore $c_{HW}$ and $c_{HB}$ should be tested in vector boson production, where their contribution relative to the SM
is quantitatively similar (indeed even $g_\rho^2/16\pi^2$ smaller) to that of $c_{W,B}$.  In this respect these effects can be practically classified as form factors.

The form factors lead to corrections to SM amplitudes whose relative size scales with energy like $E^2/m_\rho^2$.
In particular their effects on the on-shell couplings of a light Higgs are of order $m_W^2/m_\rho^2=(\gsm^2/g_\rho^2) v^2/f^2$. On the other hand the effects of the ``new couplings'' are of order  $v^2/f^2$ and thus dominate   for a strongly-coupled Higgs sector with $g_\rho^2 \gg \gsm^2$. Our conclusions differ from the widely held expectation that
 anomalous couplings between Higgs, photons and gluons should be the most important effect, given that  they   arise at one-loop level in the SM. We do find corrections to all on-shell couplings of the Higgs, including those to photons and gluons, but the origin of these corrections is the Higgs self-interaction ($c_H$) and  Higgs coupling to fermions ($c_y$) and in particular  to the top quark. The measurement of all possible Higgs production and decay channels at the 
 LHC with 300~${\rm fb}^{-1}$ should allow a test of these interactions with a sensitivity on $v^2/f^2$ of order~$0.2$.
 The detection of a deviation from the SM in this range of $v^2/f^2$, in the absence of new light states, and in particular of 
 additional light scalars, would be an indirect but clear signature of new strong dynamics involving the Higgs.
 A direct assessment of the strongly-coupled nature of the Higgs would only be obtained by observing  the self interactions
 among the Higgs and the longitudinally polarized vector bosons. The way to study
 these interactions is the same as in ordinary Higgsless theories: through the scattering of vector bosons that were collinearly radiated from elementary fermions. In the Higgsless case, the scattering amplitude among longitudinal vector boson grows like $E^2/v^2$. In our case, the light Higgs fails to fully moderate this growth and the amplitude behaves like $E^2/f^2$. As the Higgs plays the role of a fourth Goldstone, in addition to the $V_LV_L\to V_LV_L$ channels
 we also have strong double-Higgs production $V_LV_L\to hh$ with a comparable cross section. It is well known that the study of high-energy scattering among longitudinal vector bosons is not a straightforward task at the LHC. With 300~${\rm fb}^{-1}$, the expected sensitivity on
 $v^2/f^2$ is between 0.5 and 0.7. Our study further motivates analyses of $WW$ scattering, even in presence of a light Higgs.
 An interesting  issue which may be studied is the possibility to detect the reaction of double-Higgs production. Given the high energy and $p_T$ of the $b$-jets from Higgs decay and the presence of a rapidity gap, it may be possible to selects these events over the QCD background. Of course the upgraded luminosity of the second phase of LHC would make a crucial difference in these searches.

Through precise measurements of Higgs physics, the ILC will largely improve  the sensitivity on $v^2/f^2$ down to $\sim 0.01$, corresponding to $4\pi f$ up to 30~TeV. If, at this level,  no deviation from the SM is detected, it will be fair to say that the idea of a composite light Higgs is ruled out. 
Measurements of the anomalous triple gauge vertices at the ILC can test the value of $m_\rho$ up to 6--8~TeV. This sensitivity is far superior to what has been reached at LEP, through the measurement of  $\widehat S$ ($m_\rho \gsim 2$~TeV), or what can be achieved at the LHC, through direct resonance production  ($m_\rho \sim 3$~TeV).

One aspect of theories with a composite Higgs is that the top sector tends to couple with intermediate strength $\sim \sqrt {g_\rho y_t}$ to the strong sector. It is then natural to consider the possibility that one of the two helicities ($t_L$ or $t_R$) goes all the way to being composite. Moreover, in most of the realistic constructions to generate Yukawa couplings,  a  composite $t_R$ also has phenomenological advantages,
in that it allows to relax some significant bounds from $\delta \rho$ and the $Z\bar b b $ vertex. In sect.~\ref{sectop}, we have therefore extended our effective Lagrangian to the case of a fully composite $t_R$.  Although model dependent, there are important implications in flavor physics. Remarkably, for $f\sim v$ and
assuming a mixing pattern that follows the size  entries of the CKM matrix, we predict flavor effects ($B\to X_s \ell^+\ell^-$, $K^+\to \pi^+\nu\nu$, $t\to c Z$, $\Delta m_D ,\,\dots$) possibly within future experimental reach. However, the leading signature of top compositeness is associated to the reaction of four top-quark production. We plan to perform a detailed study of this and other implications of composite Higgs and top 
at the LHC in a future work. 
 
{\bf Acknowledgments} 

\noindent We would like to thank I. Antoniadis, A. Ceccucci, T. Han, B. Mele, G. Polesello, M. Porrati and especially R. Contino and I. Low for valuable conversations. This work has been partly supported by the European Commission 
under contract MRTN-CT-2004-503369.
The work of  A.P. was  partly supported   by the
FEDER  Research Project FPA2005-02211
and DURSI Research Project SGR2005-00916.

\section*{Appendix A: Integrating out vectors and scalars}

Here we describe the low-energy action obtained by integrating out fields of mass $m_\rho$ at tree level. 
We shall need the standard CCWZ notation~\cite{ccwz}
to write terms in the Goldstone action. 
The action of $g\in \G$ on the Goldstone operator $\Pi$, defined in~\eq{Umatrix}, is given by  
\beq
 gU(\Pi) \equiv U(g(\Pi)) h(\Pi,g) ,
\eeq
where 
\beq
h=e^{i\xi^a T^a}\qquad\qquad \xi^a\equiv\xi^a(\Pi,g)
\eeq
 is an element of the unbroken subgroup $\h$. Under the group action $\Pi\to g(\Pi)$ one has then $U\to  gUh^\dagger$. If $T^A$ and $T^a$ are the broken and unbroken generators respectively, we define
\beq
 U^\dagger \partial_\mu U = i\D_\mu^A T^A +i\E_\mu^a T^a \equiv i \D_\mu + i \E_\mu
 \label{DE}
 \eeq
 with transformations under $\G$ 
 \bea 
 \D_\mu(\Pi) &\to& \D_\mu(g(\Pi))=h(\Pi,g) \D_\mu(\Pi)h(\Pi,g)^\dagger  \\
 \E_\mu(\Pi)&\to& \E_\mu(g(\Pi))=h(\Pi,g) \E_\mu(\Pi)h(\Pi,g)^\dagger -ih(\Pi,g)\partial_\mu h(\Pi,g)^\dagger .
 \eea
 Notice that for space dependent $\Pi$ configurations, $\D_\mu$ and $\E_\mu$ transform
 under a local $\h$ symmetry, in particular
 $\E_\mu$ transforms like the associated gauge field.
Massive multiplets fill  reducible representations $\Phi$ of the unbroken group $\h$. The action of the global group $\G$ is realized through the
``local'' $\h$ tranformations
\beq
\Phi\to h(\Pi,g) \Phi .
\eeq
Using the last 3 equations the most general Lagrangian for the 
non-linearly realized $\G$ can be written, just by using the rules
for a ``local'' $\h$ gauge group~\cite{ccwz}. In particular $\E_\mu$ defines the
$\h$-covariant derivative $\partial_\mu +i \E_\mu$ on the massive fields $\Phi$.

The weak gauging of  $G_{SM}$ is obtained by changing
 eq.~(\ref{DE}) to $U^\dagger (\partial_\mu+iA_\mu) U \equiv  i \bar \D_\mu(\pi,A) +i \bar \E_\mu(\pi,A)$. Indeed since we treat the gauge fields as spectators in what follows, we can, without loss of generality, gauge the full $\h\supset G_{SM}$.
Thus $\bar \E_\mu^a$ transforms as the $\h$-gauge field both under the global $\G$-tranformations and under the genuinely local $g\equiv h(x)$. It is useful to have the expressions of $\bar\E$ and $\bar \D$ at lowest order in $\Pi$
\bea
\bar \E_\mu&=&A_\mu +\E_\mu(\Pi,D_\nu)=A_\mu-\frac{i}{2}\Pi\overleftrightarrow D_\mu\Pi
+\mathcal{O}(\Pi^4)\label{ebar}\\
\bar \D_\mu&=&\D_\mu(\Pi,D_\nu)\phantom{A_\mu+}=D_\mu\Pi-\frac{1}{6}\left [\Pi,\Pi\overleftrightarrow D_\mu\Pi\right ]+\mathcal{O}(\Pi^5) ,
\label{dbar}
\eea
where $D_\mu=\partial_\mu+iA_\mu$ is the $\h$ covariant derivative.
(In the last second equality of both equations we have specified to the interesting case in which $\G/\h$ is a symmetric space.)

We want to classify the 4-derivative structures that can lead to couplings involving
two Goldstones and two gauge fields.
There are 2 relevant structures
\beq
 O_1= {\rm Tr}[F_{\mu\nu}(\bar \E)F^{\mu\nu}(\bar \E)],\qquad\qquad
 O_2={\rm Tr}[\bar \D_\mu\bar \D_\nu F^{\mu\nu}(\bar \E)],
 \label{struct}
\eeq
where $F_{\mu\nu}(\bar \E)=\partial_\mu \bar\E_\nu-\partial_\nu \bar\E_\mu+i[\bar \E_\mu,\bar \E_\nu]$. 
Substituting eqs.~(\ref{ebar})--(\ref{dbar}) into~\eq{struct}
we find that $O_W$ and  $O_B$ emerge by expanding  $O_1$,
while  $O_{HW}$ and $O_{HB}$ emerge from $O_2$. It is also evident that operators of the above form cannot involve two gluons and two Higgses.
Indicating by $\bar D_\mu\equiv
\partial_\mu +i\bar\E_\mu$ the full $\h$-covariant derivative, one could write down
other structures like $\bar \D_\mu \bar D_\nu\bar D^\nu \bar \D^\nu$, 
$\bar \D_\mu \bar D^\mu\bar D_\nu \bar \D^\nu$ or $\bar \D_\mu \bar \D^\mu\bar \D_\nu \bar \D^\nu$. These are however shown to give either the same effects at dimension 6 or terms involving at least four Goldstones.

The question remains onto which effects can be generated at tree level in 
minimally coupled theories, such as Holographic Goldstones or Little Higgses.
One distinctive feature of $O_{HW}$ and $O_{HB}$ is that they give rise to
interactions involving on-shell photons and electrically neutral states. This cannot
occur at tree level in a minimally coupled theory where photon interactions
are purely dictated by covariant derivatives. On the other hand $O_W$ and $O_B$ do not lead to any extra interactions for on-shell photons, so that one
may expect them to arise at tree level by integrating out heavy states.
This is indeed the case for both Holographic Goldstones and Little Higgses, which are known to give rise to a contribution to $\widehat S\propto c_W+c_B$ through
the exchange of heavy vector states. Let us briefly outline how this effects come about within our formalism by focussing on the case of a massive vector $V$
transforming in the adjoint of $\h$. In the ungauged limit we have the option 
to choose $V_\mu$ to transform like $\E_\mu$ under $\G$. Then the most general two derivative $\G$-invariant action is given by
 \beq
 m_\rho^4{\cal L}^{0}= m_\rho^2 \D_\mu^A\D_A^\mu -\frac{1}{4} (F_{\mu\nu}^V)^2 +\frac{1}{2}m_\rho^2(V_\mu-\E_\mu)^2 ,
 \label{heavyvector}
 \eeq
 where $F_{\mu\nu}^V=\partial_\mu V_\nu-\partial_\nu V_\mu+i[V_\mu,V_\nu]$. Notice that terms involving covariant derivatives acting on the homogeneously transforming combination $\hat V_\mu \equiv V_\mu-\E_\mu$ have  more than two derivatives. 
A limitation to two derivatives,  also automatically eliminates the term $(\partial_ \mu \hat V^\mu-i[V^\mu,\hat V^\mu])^2$
which would imply the presence of a scalar  ghost with mass $\sim m_\rho$.
The gauging of $\h$ then amounts to $\E\to \bar \E$ in the above equation. According to this change $V$ will also have to transform under the genuinely local $g=h(x)$ precisely like $A$. Notice that the structure we have thus outlined
is the same of Little Higgs model with a product group structure with $V$ playing the role of the vector boson of the second $SU(2)'$, the one under which the SM
fermions are uncharged. The mass term in eq.~(\ref{heavyvector}) mimics the effects
of the Goldstone which breaks the gauge group to the, low energy, diagonal $SU(2)$, under which both $A$ and $V$ transform thus like gauge fields.
 By integrating out $V$ we find the following correction to the low-energy effective action
 \beq
 \Delta {\cal L}=-\frac{1}{4g_\rho^2}F_{\mu\nu}(\bar \E)F^{\mu\nu}(\bar \E)
 -\frac{1}{2g_\rho^2} D^\mu F_{\mu\nu}(\bar \E)  \frac{1}{\partial^2+m_\rho^2} D^\rho F_\rho^\nu(\bar \E)
 +\dots
 \label{intvectorI}
 \eeq
 where the dots indicate terms more than quadratic in the field strength and with at least 4-derivatives.  As we already explained the first term gives rise to $O_W$ and $O_B$. The second term, instead, gives rise to four derivative corrections to pure gauge kinetic terms
   \beq
   O_{2W}=\frac{1}{2g_\rho^2m_\rho^2}(D^\mu W^i_{\mu\nu})(D_\rho W^{i\rho\nu})\qquad\qquad 
   O_{2B}=\frac{1}{2g_\rho^2m_\rho^2}(\partial^\mu B_{\mu\nu})(\partial_\rho B^{\rho\nu}).
 \label{intvector2}
 \eeq

Let us consider now the effect of integrating out massive scalars. 
At the two derivative level we can have mixings of the type
$D_\mu\Phi\D^\mu$ and $\Phi \D_\mu\D^\mu$, where we have suppressed the indices; we assume, of course that the $\h$-quantum numbers of $\Phi$ and the contractions ensure $\h$-invariance.
Integrating out $\Phi$, the leading operators involve then at least 4 derivatives. At dimension 6, one is then easily convinced that only one
term is generated in general
\beq
\frac{1}{m_\rho^2}(D^2H^\dagger )(D^2 H) ,
\eeq
which can be induced by the exchange of massive doublets.
The other possible contraction at dimension 6 order, $(D_\mu D_\nu H^\dagger)(D^\nu D^\mu H)$, cannot arise from scalar exhange as it involves a $J=2$
part. The more interesting terms arise at dimension 8, and are irrelevant to our analysis.

\section*{Appendix B: Effective Lagrangian in the canonical basis}

The first term of the effective Lagrangian~(\ref{leff}) involves  Higgs derivative  terms
 that makes difficult to 
read off physical effects.
At order $\xi$,
we can eliminate these derivative terms by 
performing  the following  non-linear  redefinition of the Higgs:
\begin{equation}
h \to h - \frac{c_H \xi}{2} \left(h +  \frac{h^2}{v} + \frac{h^3}{3v^2} \right)\, .
\end{equation}
After this redefinition, the corrections to the SM Lagrangian  are now given by  
 \bea
&&{\cal L}_h 
= \xi\, \left\{ 
- \frac{m_H^2}{2v} \left[ \left(c_6-\frac {3 c_H}{2}\right) h^3 + \left( 6 c_6- \frac{25 c_H}{3}\right) \frac{h^4}{4v}+\dots \right] 
\right. \nonumber \\  
&&
+ \frac{m_f}{v} {\bar f}f \left[ \left( c_y+\frac{c_H}{2}\right) h +\left(3 c_y + c_H\right) \frac{h^2}{2v}+\dots \right]
\nonumber \\  
&&
- c_H m_W^2 \left( \frac{h}{v} +  \frac{2h^2}{v^2} +\ldots\right) W_\mu^+ W^{-\,\mu}
- \frac{c_H m_Z^2}{2} \left( \frac{h}{v} +  \frac{2h^2}{v^2} + \ldots \right) Z_\mu Z^{\mu}
\nonumber \\  
&&
+ \left( \frac{h}{v}+\frac{h^2}{2v^2}\right) \left[ \frac{g^2}{2g_\rho^2} 
\left( {\hat c}_W W^-_\mu {\cal D}^{\mu \nu} W^+_\nu +{\rm h.c.} \right)
+\frac{g^2}{2g_\rho^2} Z_\mu {\cal D}^{\mu \nu} \left[ {\hat c}_Z Z_\nu +
\left( \frac{2 {\hat c}_W}{\sin 2\theta_W}-\frac{{\hat c}_Z}{\tan\theta_W}\right) A_\nu \right]
\right. \nonumber \\ 
&&
 -\frac{ g^2}{(4\pi)^2} \left( \frac{c_{HW}}{2}
W^{+\mu \nu} W^-_{\mu \nu} +\frac{c_{HW}+\tan^2\theta_W c_{HB}}{4} Z^{\mu \nu}Z_{\mu \nu} -2 \sin^2\theta_W c_{\gamma Z}F^{\mu \nu}Z_{\mu \nu} \right) + \dots 
\nonumber \\ 
&&\left. \left. 
+\frac{\alpha g^2 c_\gamma}{4\pi g_\rho^2} F^{\mu \nu}F_{\mu \nu} +\frac{\alpha_s y_t^2 c_g}{4\pi g_\rho^2} G^{a \mu \nu}G^a_{\mu\nu}
\right] \right \}\, .
\label{lsilh2}
\eea
Eq.~(\ref{lsilh2})  could also have been  obtained  from \eq{leff}  by using the 
Higgs equation of motion.
The Lagrangian \eq{lsilh2} is in a  more suitable  basis to compare it with 
 the effective Lagrangian arising from specific  SILH models, and  extract the 
 predictions of these models for the coefficients $c_i$.
For this purpose it will be useful the give the Higgs  couplings 
corrected by  \eq{lsilh2} at  order $\xi$  and at zero momentum (or, equivalently, neglecting
$g^2\xi/g_\rho^2$ corrections). 
For the  $hWW$, $hff$ and $h^3$  coupling, we have
\bea
\displaystyle g_{hWW}  &=&  g m_W \left[ 1 -  \frac{c_H}{2} \xi \right]\, ,\label{hww}\\
\displaystyle g_{hff} & =  &\frac{g m_f}{2 m_W}
 \left[ 1 - \xi \left(\frac{c_H}{2}+c_y\right) \right]\, , \label{hff}\\ 
 \displaystyle g_{hhh} & =&  \frac{g m_H^2}{4 m_W} \left[ 1+ \xi\left(c_6 - \frac{3 c_H}{2}\right) \right]\, .\label{hhh}
\eea
%

\section*{Appendix C: Loop functions for the Higgs radiative decays}

Here we give the loop functions describing the Higgs radiative decays in eqs.~(\ref{gam4})--(\ref{gam6}),
including the $\mathcal{O}(\alpha_S)$ corrections coming from matching the SM contribution to the operator basis, in the limit of heavy top.
\beq
I_g = \frac{1}{2} F_{1/2} (x_{tH})\left( 1+\frac{11\alpha_S}{4\pi}\right) ,~~~~
I_\gamma =   \frac{4}{3} F_{1/2} (x_{tH})\left( 1-\frac{\alpha_S}{\pi}\right) ,~~~~J_\gamma = F_1 (x_{WH})
\eeq
\beq
I_Z = \frac{2(3-5t_W^2)}{3t_W}
\left[ I_1 (x_{tH},x_{tZ})- I_2 (x_{tH},x_{tZ}) \right] \left( 1-\frac{\alpha_S}{\pi} \right)
\eeq
\beq
J_Z = \left[ \left( 1+\frac{2}{x_{WH}}\right) t_W -\left( 5+\frac{2}{x_{WH}}\right) \frac{1}{t_W} \right] I_1 (x_{WH},x_{WZ}) +4\left( \frac{3}{t_W}-t_W \right)  I_2 (x_{WH},x_{WZ}) 
\eeq
\beq
t_W \equiv \tan\theta_W,~~~~~~x_{ij}\equiv \frac{4m_i^2}{m_j^2}, ~~~~ i=t,W,~~j=H,Z
\eeq
\beq
F_{1/2}(x) = -2 x \left[ 1+(1-x) f(x)\right]  ,~~~~F_1(x)=2+3x \left[ 1+(2-x) f(x)\right] 
\eeq
\beq
I_1 (x,y) =\frac{xy}{2(x-y)}\left\{ 1+\frac{xy}{x-y} \left[ f(x)-f(y)\right] +\frac{2}{x-y}\left[ g(x)-g(y)\right] \right\}
\eeq
\beq
I_2 (x,y) =-\frac{xy}{2(x-y)}\left[ f(x)-f(y)\right]
\eeq
\beq
f(x)=\arcsin^2 \left( x^{-1/2}\right) ,~~~~g(x)=\sqrt{x-1}\arcsin \left( x^{-1/2}\right) .
\eeq


\end{document}